# Elasto-plastic and adhesive contact: an improved linear model and its application

*Wenguang Nan[1, 2*], Wei Pin Goh[2], Mohammad Tarequr Rahman[1]*

*1. School of Mechanical and Power Engineering, Nanjing Tech University, Nanjing 211816, China*

*2. School of Chemical and Process Engineering, University of Leeds, Leeds LS2 9JT, UK*

**Contact Email: nanwg@njtech.edu.cn*

**Abstract:** An improved linear model is developed for elasto-plastic and adhesive contact. New correlations are proposed and validated to estimate the key input parameters of the model, including contact stiffness, yield point, maximum pull-off force and time step. The newly proposed contact model is applied to the analysis of single particle contact behaviour upon impact and bulk particle flow behaviour by DEM simulations. The results show that both single particle and bulk powder behave more "cohesively" if contact plastic deformation is taken into consideration. A cohesion yield number is proposed to describe and govern the extent of yielding when cohesive particles are in contact with each other. There is critical particle size, below which the effect of plastic deformation must be considered. This provides a new framework and criteria for elasto-plastic and adhesive contact model, and a step towards understanding the effect of plastic deformation on the behaviour of cohesive particles.

## 1 Introduction

The macroscopic bulk behaviour of powders is governed by the microscopic activities of the individual particles in a granular assembly. Predicting the bulk behaviour of a particulate system requires a thorough understanding of the dynamics of individual particles at microscopic level, but this is very difficult to achieve experimentally. As an alternative method, numerical simulations by Discrete Element Method (DEM) (Cundall and Strack, 1979) are usually used. In DEM simulation, the particle interaction is described by a contact model of choice and the particle motion is governed by Newton's law. To accurately represent a realistic particulate system, the particle physical and mechanical properties (e.g. size and shape distribution, Young's modulus, density) and interaction parameters (e.g. friction coefficient, restitution coefficient) should be characterised experimentally at single particle level (Nan et al., 2018; Pasha et al., 2020), or artificially tuned to produce matching results with the bulk calibration test (e.g. repose angle, uniaxial compression test) (Johnstone, 2010), and the contact model should obey realistic physical deformation law of the particle material.

Particle contact can be divided into four classes: elastic, elastic-adhesive, elastic-plastic, and elastic-plastic with adhesion, depending on the material used. For elastic contact, deformation is recoverable, and the normal contact force can be well described by Hertz contact model (Thornton, 2015). For elastic-adhesive contact, as the particles tend to stick to each other, additional energy is required to separate them. The normal contact force in the elastic-adhesive contact can be predicted by JKR theory (Johnson et al., 1971). JKR theory extends the Hertz model to the elastic-adhesive contact by using an energy balance approach, and the contact area is larger than that of Hertz model. The normal contact force in the elastic-adhesive contact could also be predicted by DMT theory (Derjaguin et al., 1975), which assumes that the adhesive force does not affect the contact area and considers the adhesive force and Hertz force separately. In DEM simulation, contact is usually assumed to be either elastic or elastic-adhesive, and the aforementioned contact models have been widely applied to simulate various particulate systems, such as coating (Pasha et al., 2016) and powder spreading (Nan et al., 2018; Nan et al., 2020). However, most materials first deform elastically, which then followed by a plastic deformation. This is especially true for rough particles with tiny asperities on the surface. In elasto-plastic contact, a portion of the particle deformation is recoverable, and the rest is permanent. To account for plastic deformation, the normal contact force could be calculated by the non-linear model of Thornton and Ning (1998), or the linear model of Walton and Braun (1986). However, in the latter, the initial elastic deformation stage is omitted, which is not realistic. If cohesive force is involved, the contact could become elasto-plastic with adhesion, and the corresponding contact model becomes more complex as the overlap and negative force at the detachment point are affected by the plastic deformation (i.e. permeant deformation with flattened area). For this specific type of contact, the non-linear model of Thornton and Ning (1998), the linear model of Pasha et al. (2014) and Luding (2008) can be used to estimate the normal contact force. In the model of Luding (2008), the contact breaks at zero overlap, which is unpragmatic since plastic deformation is permanent and hence the detachment must take place at non-zero overlap. This model assumes that all particles undergo plastic deformation. But in the reality, for dynamic particle system, the external load applied on the particles at some regions is not large enough to cause yielding and the contact is still elasto-adhesive. In the non-linear model of Thornton and Ning (1998), the equations of normal contact force is derived based on material properties and contact mechanics theory, but the equations are in very complex form, and the contact model is very computationally expensive due to its non-linearity nature. The model of Pasha et al. (2014) is a linear version of the non-linear model of Thornton and Ning (1998), and it is more attractive than the model of Luding (2008) and Thornton and Ning (1998) in terms of physical nature and computational time. However, Pasha et al. (2014) mainly proposed a framework for the contact model, and many details are not provided, such as the unified mathematical equations of the contact force and the estimation criteria for the contact parameters. Pasha et al. (2014) proposed a simplified version for further analysis of the effect of plastic deformation on the particle system, but the initial elastic process is omitted. It should be noted that in both models of Luding (2008) and Pasha et al. (2014), the adhesive sticking velocity predicted by JKR theory is not guaranteed anymore.

Based on the work of Thornton and Ning (1998) and Pasha et al. (2014), an improved linear model is proposed here for the elasto-plastic and adhesive contact, with valid criteria to estimate the input parameters involved in the model. The contact model is applied to the analysis of single particle contact behaviour upon impact and bulk particle flow behaviour in FT4 rheometer by DEM simulations. This provides a linear elasto-plastic and adhesive contact model, which is well suited for the DEM simulations, and also a step towards understanding the effect of plastic deformation on the sticking velocity of single cohesive particle and flowability of bulk cohesive powder.

## 2 Description of the contact model

The total contact force, $\boldsymbol{F}$, is the sum of normal contact force $\boldsymbol{f}_n$, tangential contact force $\boldsymbol{f}_t$, and the damping force ($\boldsymbol{f}_{nd}$, $\boldsymbol{f}_{td}$):

$$\boldsymbol{F}_n = \boldsymbol{f}_n + \boldsymbol{f}_{nd} \tag{1}$$

$$\boldsymbol{F}_t = \boldsymbol{f}_t + \boldsymbol{f}_{td} \tag{2}$$

### 2.1 Normal contact force

The normal contact force $\boldsymbol{f}_n$ is given as:

$$\boldsymbol{f}_n = f\boldsymbol{n} \tag{3}$$

where $\boldsymbol{n}$ is the unit vector in the normal direction; $f$ is the magnitude of normal contact force, which is linear to the normal overlap $\alpha$, as shown in Fig. 1.

If the maximum load applied is less than the yield force $f_y$ of the particle, there will be no plastic deformation at the end of loading process, thus, the contact behaves elastically with adhesion. In this case, Fig. 1 could be simplified to Fig. 2(a)-(b), where the contact is considered as a linear version of JKR model (Thornton, 2015). During the loading stage, the normal contact force $f$ suddenly drops to -$f_0$ when a contact is established, $f$ then increases linearly with the normal overlap $\alpha$, the slope of which is dictated by the elastic stiffness $k_e$. During the unloading stage, the contact force is non-zero even for negative overlap, as further work is required to separate the cohesive contact. The contact breaks at a negative overlap $\alpha_{fe}$ with contact force of $-5f_{ce}/9$.

If the maximum load is larger than the yield force $f_y$, there will be a plastically-deformed domain inside the contact area, resulting in plastic deformation before the end of the loading process. At the loading stage, as shown in Fig. 2(c), $f$ increases linearly with the normal overlap, the slope of which follows that of plastic stiffness $k_p$ in the plastic phase, where $k_p$ is usually less than $k_e$. At the unloading stage, with a decrease in normal overlap, $f$ initially decreases linearly with the elastic stiffness $k_e$ until it reaches point $E$ ($\alpha$=$\alpha_{cp}$, $f$=-$f_{cp}$), where a maximum pull–off force $f_{cp}$ is obtained. $f$ then increases slowly with a stiffness of $k_c$ until it reaches point $F$ ($\alpha$=$\alpha_{fp}$, $f$=-5$f_{cp}$/9), where the particles are detached. If a normal overlap is still identified after the detachment, i.e. the particle centre distance is less than the sum of contact radius, the plastic deformation is maintained. Thus, during the reloading stage, the contact could only be re-established at $\alpha$=$\alpha_{c0}$ with an initial value of -$f_{0p}$ (i.e. -8$f_{cp}$/9), as shown in Fig. 2(d). With an increase in normal overlap at the reloading stage, the contact initially behaves elasto-adhesively with a stiffness of $k_e$ until $f$ reaches point $D$ ($\alpha$=$\alpha_{\max}$), where the maximum normal force in previous loading stage is reached, and then plastic deformation prevails with a plastic stiffness of $k_p$.

Fig. 1. Schematic diagram of the normal force $f$-overlap $\alpha$ relationship in the improved linear elasto-plastic and adhesive model.

Fig. 2. Schematic diagram of the normal force $f$-overlap $\alpha$ relationship in the loading/unloading and reloading processes for the contact: (a-b)-without plastic deformation ($f_{\max}$<$f_y$); (c-d)-with plastic deformation ($f_{\max}$>$f_y$).

For the contact before yielding (i.e. line BC), $k_e$ is a constant and does not vary with normal overlap. However, after the yielding of the contact, $k_e$ would increase with the normal overlap from which unloading commences, the details of which will be discussed in Section 3.3. Thus, the elastic stiffness $k_e$ of line BC in Fig. 1 corresponds to the lower limit of $k_e$, and it is assigned as $k_{el}$ for convenience.

Similar to JKR model, $f_0$ at $\alpha$=0 in Fig. 1 is given as:

$$f_0 = \frac{8}{9} f_{ce} \tag{4}$$

$$f_{ce} = 1.5\pi \Gamma R^* \tag{5}$$

where $f_{ce}$ is the maximum pull-off force before yielding; $\Gamma$ is the surface energy; $R^*$ is the equivalent radius. The critical normal overlaps in Fig. 1 are given as:

$$\alpha_0 = \frac{f_0}{k_{el}} = \frac{8}{9}\frac{f_{ce}}{k_{el}} \tag{6}$$

$$\alpha_y = \alpha_0 + \frac{f_y}{k_{el}} \tag{7}$$

$$\alpha_{ce} = \alpha_0 - \frac{f_{ce}}{k_{el}} \tag{8}$$

$$\alpha_{fe} = \alpha_0 - \frac{f_{ce}}{k_{el}} - \frac{4}{9}\frac{f_{ce}}{k_{cl}} \tag{9}$$

$$\alpha_{\max} = \alpha_y + \frac{f_{\max} - f_y}{k_p} \text{ with } \alpha_{\max} = \max(\alpha_{\max}, \alpha_y) \tag{10}$$

$$\alpha_p = (1 - \frac{k_p}{k_e})(\alpha_{\max} - \alpha_y) + (1 - \frac{k_{el}}{k_e})(\alpha_y - \alpha_0) + \alpha_0 \tag{11}$$

$$\alpha_{c0} = \alpha_p - \frac{8}{9}\frac{f_{cp}}{k_e} \tag{12}$$

$$\alpha_{cp} = \alpha_p - \frac{f_{cp}}{k_e} \tag{13}$$

$$\alpha_{fp} = \alpha_p - \frac{f_{cp}}{k_e} - \frac{4}{9}\frac{f_{cp}}{k_c} \tag{14}$$

where $\alpha_p$ is derived from the normal force at point $D$:

$$k_p(\alpha_{\max} - \alpha_y) + f_y = k_e(\alpha_{\max} - \alpha_p) \tag{15}$$

$$\alpha_p = (1 - \frac{k_p}{k_e})\alpha_{\max} + \frac{k_p\alpha_y - f_y}{k_e} \tag{16}$$

For the contact before yielding, $\alpha_{\max}$ in Eq. (10) is mathematically reduced to $\alpha_{\max}=\alpha_y$ while $k_e$ is reduced to its minimum value (i.e. $k_{el}$), thus, Eq. (11) is reduced to $\alpha_p=\alpha_0$, and $f_{cp}$ is reduced to $f_{cp}=f_{ce}$ (shown in Section 3.4), $k_c$ is reduced to $k_{cl}$ (shown in Section 3.2), resulting in $\alpha_{fp}=\alpha_{fe}$, $\alpha_{cp}=\alpha_{ce}$, and $\alpha_{c0}=0$. Therefore, Eqs. (11)-(14) could be deemed as a general mathematical form for the calculation of critical normal overlaps, which are valid for both the contacts before and after yielding.

The non-zero normal force could be classified into three states, $f_c$ for line AB or EF, $f_e$ for line BC

or DE, $f_p$ for line CD:

$$f_e = k_e(\alpha - \alpha_p) \tag{17}$$

$$f_p = f_y + k_p(\alpha - \alpha_y) \tag{18}$$

$$f_c = -f_{cp} + k_c(\alpha_{cp} - \alpha) \tag{19}$$

Thus, the normal force $f$ in Fig. 1 is mathematically given as:

$$f = \begin{cases} f_e & \alpha > \alpha_{cp} \ \& \ f_p > f_e \\ f_p & \alpha > \alpha_{cp} \ \& \ f_p \le f_e \\ f_c & \alpha_{fp} \le \alpha \le \alpha_{cp} \\ 0 & \alpha < \alpha_{fp} \ || \ cs = 0 \end{cases} \tag{20}$$

where "$cs$=0" refers to the states: first loading with $\alpha$<0 (Fig. 2(a)) and reloading with $\alpha<\alpha_{c0}$ (Fig. 2(d)). It should be noted that Eqs. (17)-(20) are valid for both the contacts before and after yielding. For example, for the contact before yielding, as shown above, $k_e=k_{el}$, $\alpha_p=\alpha_0$, thus, Eq. (17) is simplified to $f_e=k_{el}(\alpha- \alpha_0)$, which is intuitively expected in Fig. 1.

### 2.2 Tangential contact force

The tangential contact force $\boldsymbol{f}_t$ is given as:

$$\boldsymbol{f}_t = k_t \boldsymbol{\delta}_t \tag{21}$$

where $\boldsymbol{\delta}_t$ is the vector of tangential displacement; $k_t$ is the tangential stiffness, which is related to $k_n$, given as:

$$\frac{k_t}{k_n} = 4\frac{G^*}{E^*} \tag{22}$$

where $G$* and $E$* are the equivalent shear modulus and Young's modulus, respectively; $k_n$ is the normal stiffness, i.e. $k_c$ for line AB or EF, $k_e$ for line BC or DE, $k_p$ for line CD, as shown in Fig. 1. For the sliding contact, i.e. $k_t|\boldsymbol{\delta}_t|>\mu f$, the energy is dissipated from the interfacial sliding without introducing the viscous damping in the tangential direction. Thus, Eq. (2) is reduced to $\boldsymbol{F}_t=\boldsymbol{f}_t$, given as:

$$\boldsymbol{F}_t = \mu f \frac{\boldsymbol{\delta}_t}{|\boldsymbol{\delta}_t|} \tag{23}$$

where $\mu$ is the sliding friction coefficient; $f$ is normal force in Eq. (20).

## 2.3 Damping force

For the contact before yielding, besides the frictional dissipation through interfacial sliding and adhesive work, the energy dissipation is mainly attributed to the viscous elastic damping, especially in the normal direction. The damping force in the normal and tangential direction is given as:

$$f_{nd} = 2\gamma\sqrt{m^* k_n} V_n \tag{24}$$

$$f_{td} = 2\gamma\sqrt{m^* k_t} V_t \tag{25}$$

where $\gamma$ is the damping coefficient due to viscous and viscoelastic damping effect or energy dissipation of elastic wave propagation; $m^*$ is the equivalent mass; $\boldsymbol{V}_n$ and $\boldsymbol{V}_t$ are the relative velocity in normal and tangential direction, respectively. $\gamma$ is related to elastic restitution coefficient $e_0$, given as:

$$\gamma = -\beta \frac{\ln e_0}{\sqrt{\pi^2 + (\ln e_0)^2}} \tag{26}$$

where $\beta$=1 is the damping factor.

For the contact after yielding, the energy dissipation of elastic wave propagation during impact is very small, compared to the ones due to plastic deformation, as reported by Ning et al. (1995). However, the viscous damping is necessary to suppress the oscillation. For example, in the impact test, if the viscous damping force is omitted while the particle could not rebound, the contact force oscillates indefinitely along line D-E-F (Fig. 1), and a state of equilibrium can never be achieved as no energy is dissipated, as shown in Fig. 3. Thus, a small value of $\beta$ less than 1 should be used, i.e. 0.1.

Fig. 3. Oscillation of normal contact force with time if there is no viscous damping after yielding.

## 3. Key parameters of the contact model

In the improved contact model, the key parameters, i.e. yield point ($f_y$, $\alpha_y$), stiffness ($k_e$, $k_c$, $k_p$) and maximum pull-off force ($f_{cp}$), are related to the work of deformation shown in Fig. 4, and they are discussed in this section.

Fig. 4. Work of deformation due to normal contact force in different contact stages.

### 3.1 Contact force and normal overlap at yield point, $f_y$ and $\alpha_y$

The contact force and normal overlap at the yield point (point C in Fig. 1) are evaluated by assuming the same yield work as predicted by Thornton and Ning (1998), given as:

$$W_2 = \frac{f_y^2}{2k_{el}} = W_y = \frac{(\pi R^* p_y)^5}{60E^{*4} R^{*2}} \tag{27}$$

Thus, the yield force and the corresponding normal overlap are given as:

$$f_y = f_{y0}\sqrt{\frac{6}{5}\frac{k_{el}}{\pi R^* p_y}} \tag{28}$$

$$\alpha_y - \alpha_0 = \frac{f_y}{k_{el}} = \alpha_{y0}\sqrt{\frac{8}{15}\frac{\pi R^* p_y}{k_{el}}} \tag{29}$$

where $f_{y0}$ and $\alpha_{y0}$ are the yield force and the corresponding normal overlap predicted by Thornton and Ning (1998) in their non-linear contact model:

$$f_{y0} = \frac{\pi^3 R^{*2} p_y^3}{6E^{*2}} \tag{30}$$

$$\alpha_{y0} = \frac{\pi^2 R^* p_y^2}{4E^{*2}} \tag{31}$$

where $p_y$ is the limiting contact pressure of the softer particle, i.e. $p_y$=min($p_{y1}, p_{y2}$) . $p_y$ is related to the yield stress $\sigma_y$, given by Jackson & Green (2005):

$$p_y = C\sigma_y \tag{32}$$

where $C$ is a coefficient related to its Poisson's coefficient $v$, given by $C$=1.295exp(0.736$v$). It is similar to the model given by Chang et al. (1987):

$$p_y = KH \tag{33}$$

where $K$ is the hardness factor, given by $K$=0.454+0.41$v$; and $H$ is the hardness, given by $H$=2.8$\sigma_y$. In particulate systems, i.e. $v$=0.2~0.4, both models give almost the same ratio of the yield contact pressure to yield stress, $p_y/\sigma_y$=1.4~1.7, as shown in Fig. 5.

For most granular system, the yield stress can be evaluated from the quasi-static test, i.g. nano-indentation test, the data of which can be referred to the material database or handbook. However, for the particulate systems with high collision velocity, such as impact breakage, the yield stress is dynamic and sensitive to the strain rate (Burgoyne and Daraio, 2014; Ning, 1995). In this case, the yield stress should be experimentally evaluated from the dynamic test (e.g. dropping hammer experiment), which could be fitted by the following model:

$$\sigma_y = \sigma_y^s (1 + Y \ln(V_i / V_0)) \tag{34}$$

where $V_i$ is the characteristic strain rate of the particulate system in the simulation; $\sigma_y$ is the characteristic yield stress at $V_i$; $V_0$ is the quasi-static strain rate at which $\sigma_y^s$ is measured; $Y$ is an empirical parameter determined experimentally by the dynamic test.

Fig. 5. Variation of the ratio of yield contact pressure $p_y$ to yield stress $\sigma_y$ with Poisson's ratio.

### 3.2 Stiffness $k_c$

As shown in Fig. 2(b), for the contact without yielding, the adhesive sticking work $W_0$ needs to be overcome to separate the particles in contact. $W_0$ is assumed to be the same as traditional JKR model, given as:

$$W_0 = \frac{56}{162}\frac{f_{ce}^2}{k_{cl}} + \frac{17}{162}\frac{f_{ce}^2}{k_{el}} = W_{JKR} \tag{35}$$

where the derivation of $W_0$ is shown in Appendix A; $k_{el}$ and $k_{cl}$ are the stiffnesses of the contact before yielding, corresponding to the lower limit of $k_e$ and $k_c$, respectively. The ratio of $k_{cl}$ to $k_{el}$ is given as:

$$\frac{k_{cl}}{k_{el}} = \frac{\frac{56}{162}\frac{f_{ce}^2}{k_{el}}}{W_{JKR} - \frac{17}{162}\frac{f_{ce}^2}{k_{el}}} \tag{36}$$

where $W_{JKR}$ is given as (Thornton, 2015):

$$W_{JKR} = 7.09\left(\frac{\Gamma^5 R^{*4}}{E^{*2}}\right)^{1/3} \tag{37}$$

Substituting Eqs. (5) and (37) into Eq. (36), given as:

$$\frac{k_{el}}{k_{cl}} = 0.92\frac{k_{el}}{\Gamma^{1/3} E^{*2/3} R^{*2/3}} - 0.3 \tag{38}$$

In most granular systems, $k_{el}/k_{cl}$ predicted by Eq. (38) is larger than 1. If comparing $k_{cl}$ to the normal stiffness at $\alpha$=0 predicted by Hertz model with JKR theory, it is given as:

$$\frac{k_{el}}{k_{cl}} = 1.13\frac{k_{el}}{k_{H-JKR,\,\alpha=0}} - 0.3 \tag{39}$$

$$k_{H-JKR,\,\alpha=0} = 1.23(\Gamma E^{*2} R^{*2})^{1/3} \tag{40}$$

where the derivation of $k_{H\text{-}JKR,\ \alpha=0}$ is shown in Appendix B. Thus, the stiffness $k_{cl}$ in Eq. (38) is very close to the stiffness at $\alpha$=0 predicted by Hertz model with JKR theory.

For the contact after yielding (Fig. 2(c)-(d)), although $k_e$ in the unloading process increases with the maximum load, the ratio of $k_e/k_c$ is assumed to be the same as $k_{el}/k_{cl}$ predicted by Eq. (38).

### 3.3 Stiffness $k_e$

In the unloading process after yielding, i.e. line *DE* in Fig. 1, the elastic stiffness $k_e$ should vary with $\alpha_{\max}$, as reported by Luding (2008), Pasha et al. (2014), Thornton et al. (2017). In the work of Luding (2008) and Pasha et al. (2014), $k_e$ varies linearly with $\alpha_{\max}$. However, as reported by Thornton et al. (2017), $k_e$ should be a linear function of $\sqrt{\alpha_{\max}}$ , which is also in accordance with Hertz model:

$$k_H = \frac{df_H}{d\alpha} = 2E^*R^*\sqrt{\alpha / R^*} \tag{41}$$

where $f_H$=4/3$E^*R^{*1/2}\alpha^{3/2}$ is the elastic force in Hertz model. Here, in the linear elasto-plastic and adhesive contact model developed in this wrok, $k_e$ is given as:

$$k_e = k_{em}\sqrt{\frac{\alpha_{\max}}{R^*}} \tag{42}$$

where $k_{em}$ is assumed to be the maximum stiffness at a normal overlap of $R^*$.

The elastic stiffness $k_e$ in Eq. (42) can be evaluated through the given unloading line after yielding, i.e. $k_{e0}$ at $\alpha_{\max 0}$ is known, given as:

$$k_{em} = k_{e0}\sqrt{\frac{R^*}{\alpha_{\max 0}}} \tag{43}$$

Thus, $k_e$ for line *DE* is given as:

$$k_e = k_{e0}\sqrt{\frac{\alpha_{\max}}{\alpha_{\max 0}}} \tag{44}$$

For the contact before yielding, i.e. $\alpha<\alpha_y$ for line *BC*, $k_{el}$ is the lower limit of $k_e$, and it is calculated by substituting $\alpha_{max}$= $\alpha_y$ into Eq. (44), given as:

$$k_{el} = k_e(\alpha_{max} = \alpha_y) = k_{e0}\sqrt{\frac{\alpha_y}{\alpha_{max0}}} \tag{45}$$

As $\alpha_y$ is a function of $k_{el}$, as shown in Eq. (29), Eq. (45) should be solved together with Eq. (29) in an iterative fashion by initially assuming $\alpha_y$=$\alpha_{y0}$.

The elastic stiffness $k_e$ in Eq. (42) can also be evaluated through the stiffness of the contact before yielding, i.e. $k_{el}$ for line *BC* in Fig. 1 is known. In this case, $\alpha_y$ can be directly calculated from Eq. (29). Similar to Eq. (44), the elastic stiffness $k_e$ for line *DE* is given as:

$$k_e = k_{el}\sqrt{\frac{\alpha_{max}}{\alpha_y}} \tag{46}$$

In the work of Luding (Luding, 2008) and Pasha et al. (Pasha et al., 2014), to avoid $k_e$ being less than $k_p$ at small plastic deformation, $k_p$ is included into the calculation of $k_e$. If applying this concept to Eq. (42), $k_e$ is then given as:

$$k_e = k_p + (k_{em} - k_p)\sqrt{\frac{\alpha_{max}}{R^*}} \tag{47}$$

$$k_e = k_p + (k_{e0} - k_p)\sqrt{\frac{\alpha_{max}}{\alpha_{max0}}} \tag{48}$$

$$k_{el} = k_p + (k_{e0} - k_p)\sqrt{\frac{\alpha_y}{\alpha_{max0}}} \tag{49}$$

To validate these two methods (i.e. without and with involving $k_p$ during the calculation of $k_e$ ), the force-overlap response of loading/unloading predicted by Ning (1995) and Du et al. (2007) is used. In the work of Ning (1995), the impact test of an ammonium fluorescein particle to a silicon target at three impact velocity (i.e. 2, 5 and 10 m/s) is analysed by using the non-linear elasto-plastic and adhesive model developed by Thornton and Ning (1998), while in the work of Du et al. (2007), the indentation test of a ruthenium particle to a rigid and flat surface is analysed by using finite element model. The particle properties used in their work are shown in Table 1, and the equivalent Young's modulus is given by $E^*=E/(1-v^2)$ as the wall in their work has a much larger Young's modulus than that of the particle. By using the Digitizer Tool of Origin software (Originlab, USA), $f_{max}$, $\alpha_{max}$, $\alpha_p$ and $f_{cp}$ are digitized from the three unloading curves in their work, as shown in Table 2, and $k_e$ and $k_p$ are re-calculated as:

$$k_{e,i} = \frac{f_{\max,i}}{\alpha_{\max,i} - \alpha_{p,i}} \quad (i = 1-3) \tag{50}$$

$$k_p = \frac{1}{2}\left(\frac{f_{\max,1} - f_{\max,2}}{\alpha_{\max,1} - \alpha_{\max,2}} + \frac{f_{\max,2} - f_{\max,3}}{\alpha_{\max,2} - \alpha_{\max,3}}\right) \tag{51}$$

By specifying the maximum value of $\alpha_{\max}$ and the corresponding elastic stiffness $k_e$ in Table 2 as $\alpha_{\max0}$ and $k_{e0}$, respectively, the variation of $k_e$ with normal overlap can be predicted by Eq. (44) or Eq. (48), as shown in Fig. 6. The prediction of Hertz model is also included in Fig. 6, which is the same as that of JKR theory at the same normal overlap for the unloading process with plastic deformation (i.e. $\alpha >> \alpha_0$), as shown in Appendix B.

In Fig. 6(a), the prediction of Eqs. (42)-(45) agrees well with the ones re-calculated from Ning et al. (Ning, 1995), which is intuitively expected as both of them are originated from Hertz model. According to the theory of Thornton (2015), $k_e$ should be less than the value predicted by Hertz model (i.e. Eq. (41)) if cohesion and plastic deformation have significant effects on the unloading process. As shown in Fig. 6(a), $k_e$ predicted by Eqs. (42)-(45) fulfils this criterion even at small plastic deformation. On the contrary, $k_e$ is overestimated by Eqs. (47)-(49) especially at small plastic deformation, as $k_p$ contributes more to $k_e$ than that of $\alpha_{\max}^{0.5}$ at small $\alpha_{\max}$. In fact, in Eqs. (42)-(45), the minimum predicted value of $k_e$ is always larger than $k_p$, as long as the contact in the unloading process obeys the same law of Hertz theory. Thus, $k_p$ is not needed in the calculation of $k_e$ as they are actually independent parameters from the physics point of view.

In Fig. 6(b), the prediction of Eqs. (42)-(45) agrees better with the ones re-calculated from Du et al. (2007) than that of Eqs. (47)-(49). The elastic stiffness in Du et al. (2007) is very large, resulting in small effects of adhesion and plastic deformation on the unloading process. Thus, $k_e$ is very close to the ones predicted by Hertz model, which is very different to the work of Ning (1995). It should be noted that in the finite element simulation of Du et al. (2007), different assumptions are used and $k_p$ decreases significantly at small plastic deformation. Thus, if using Eqs. (47)-(49) to calculate $k_e$, a smaller and changeable $k_p$ should be used for the comparison with $k_e$ at small plastic deformation.

The contact parameters at the yield point (i.e. $\alpha=\alpha_y$) is shown in Table 3. In the theory of Thornton and Ning (1998), the plastic loading line is tangential to the elastic loading curve predicted by Hertz model, resulting in $k_e$ at the yield point being equal to $\pi R^*p_y$ while $k_p=k_e$. In this work, $k_{el}$ predicted by Eq. (45) is very close to $\pi R^*p_y$, which agrees well with this theory, whilst $k_{el}$ predicted by Eq. (49) is much over-estimated. Thus, $k_e$ in the following sections is calculated by Eqs. (42)-(45) unless otherwise specified. Meanwhile, if there are no experimental data available for the input parameters, i.e. $k_p$, and $k_{el}$ (or $\alpha_{\max 0}$ and $k_{e0}$), the following value is recommended for $k_{el}$, and $k_p$ could be assumed to be the same as $k_{el}$.

$$k_{el} = 2E^*\sqrt{R^*\alpha_{y0}} = \pi R^* p_y \tag{52}$$

Table 1. Particle properties used in the work of Ning (1995) and Du et al. (2007).

Table 2. Re-calculated particle contact parameters based on the plot in Ning (1995) and Du et al. (2007).

Fig. 6. Comparisons of stiffness $k_e$ predicted by different methods with (a) Ning (1995) and (b) Du et al. (2007) under different maximum normal overlaps $\alpha_{max}$ (i.e. the overlap at which the unloading commences).

Table 3. Particle contact parameters at the yield point ($\alpha=\alpha_y$) for the particles used in Ning (1995) and Du et al. (2007).

### 3.4 Maximum pull-off force $f_{cp}$

Compared to the contact before yielding, more adhesive sticking work is needed to separate the adhesive particles with plastic deformation. The increment is due to the flattening of the contact area, given as:

$$W_6 - W_0 = \pi a_{res}^2 \Gamma \tag{53}$$

where $a_{res}$ is the contact area at the residual deformation $\alpha_{res}$, and $\alpha_{res}^2=R^*\alpha_{res}$ is assumed. To be consistent with the reloading process shown in Fig. 2(d), where the contact could re-establish at the normal overlap $\alpha_{c0}$, the residual deformation is assumed to be $\alpha_{res}=\alpha_{c0}$. Thus, $\alpha_{res}$ could be reduced to 0 for the contact without plastic deformation, which is intuitively expected. According to Eqs. (5)-(6) and (12), the right-hand side of Eq. (53) is given as:

$$\pi {a_{res}}^2 \Gamma = \pi \Gamma R^* \alpha_{c0} = \pi \Gamma R^* (\alpha_p - \frac{8}{9}\frac{f_{cp}}{k_e}) = \frac{2}{3}(f_{ce}\alpha_p - f_{cp}\alpha_0 \frac{k_{el}}{k_e}) \tag{54}$$

According to Eqs. (A1) and (A6), the left-hand side of Eq. (53) is given as:

$$W_6 - W_0 = \frac{16}{27}\frac{1}{A}(\frac{{f_{cp}}^2}{k_e} - \frac{{f_{ce}}^2}{k_{el}}) = \frac{16}{27}\frac{f_{ce}}{A}\frac{f_{ce}}{k_{el}}(\frac{{f_{cp}}^2}{{f_{ce}}^2}\frac{k_{el}}{k_e} - 1) = \frac{2}{3}\frac{f_{ce}\alpha_0}{A}(\frac{{f_{cp}}^2}{{f_{ce}}^2}\frac{k_{el}}{k_e} - 1) \tag{55}$$

where $A$ is given as:

$$A = \frac{16}{27} / (\frac{56}{162}\frac{k_{el}}{k_{cl}} + \frac{17}{162}) = \frac{16}{27} / (\frac{56}{162}\frac{k_e}{k_c} + \frac{17}{162}) \tag{56}$$

Substituting Eqs. (54) and (55) into (53) yields:

$$\frac{k_{el}}{k_e}\frac{{f_{cp}}^2}{{f_{ce}}^2} - 1 = A(\frac{\alpha_p}{\alpha_0} - \frac{k_{el}}{k_e}\frac{f_{cp}}{f_{ce}}) \tag{57}$$

which can be further simplified as:

$$(\frac{f_{cp}}{f_{ce}})^2 + A\frac{f_{cp}}{f_{ce}} - (\frac{\alpha_p}{\alpha_0}A + 1)\frac{k_e}{k_{el}} = 0 \tag{58}$$

Thus, the maximum pull-off force $f_{cp}$ is given as:

$$\frac{f_{cp}}{f_{ce}} = \frac{-A + \sqrt{A^2 + 4\frac{k_e}{k_{el}}(\frac{\alpha_p}{\alpha_0}A + 1)}}{2} \tag{59}$$

For the contact before yielding, i.e. $\alpha_p = \alpha_0$, $k_e = k_{el}$, Eq. (59) is mathematically simplified to $f_{cp} = f_{ce}$. Thus, Eq. (59) is a general form for both contacts before and after yielding shown in Fig. 2.

In the work of Pasha et al. (2014), $f_{cp}$ is given as:

$$f_{cp} = -\sqrt{\frac{162}{137}\pi \Gamma R^* k_e (\alpha_p - \alpha_y)(2 - \frac{\alpha_p - \alpha_y}{R^*})} \tag{60}$$

which is derived for constant $k_e$. If considering the fact that $k_e$ varies with $\alpha_{\max}$, it can be further simplified by using similar derivations as above, given as:

$$\frac{f_{cp}}{f_{ce}} = \sqrt{\frac{162}{137}\frac{\pi \Gamma R^*}{f_{ce}}\frac{k_e}{f_{ce}}(\alpha_p - \alpha_y)(2 - \frac{\alpha_p - \alpha_y}{R^*})} = \sqrt{B\frac{k_e}{k_{el}}\frac{(\alpha_p - \alpha_y)}{\alpha_0}(2 - \frac{\alpha_p - \alpha_y}{R^*})} \tag{61}$$

$$B = \frac{162}{137} \times \frac{16}{27} = \frac{96}{137} = 0.7 \tag{62}$$

For the case with small plastic deformation, i.e. $\alpha_{\max}$ is slightly larger than $\alpha_y$ but with $\alpha_p < \alpha_y$, Eq. (60)

or Eq. (61) will predict zero pull-off force, which is unrealistic.

Thornton and Ning (1998) also proposed a method to calculate $f_{cp}$ in their non-linear model, which is given as:

$$\frac{f_{cp}}{f_{ce}} = \frac{R_{eff}}{R^*} \tag{63}$$

where $R_{eff}$ is the effective curvature due to contact flattening, and it is calculated from the equivalent elastic-adhesive normal force $f_{eq}$ at the same contact radius from which unloading commences. If applying this method to the linear model in this work, it is given as:

$$\frac{f_{cp}}{f_{ce}} = \frac{R_{eff}}{R^*} = \frac{f_{eq}}{f_{\max}} = \frac{f_y + k_e(\alpha_{\max} - \alpha_y)}{f_y + k_p(\alpha_{\max} - \alpha_y)} \tag{64}$$

For large loading force (i.e. large $\alpha_{\max}$), $f_{cp}/f_{ce}$ approaches to $k_e/k_p$.

Eq. (37) can also be used to estimate the adhesive work for the contact with plastic deformation by replacing $R^*$ with $R_{eff}$, given as:

$$\frac{W_6}{W_0} = (\frac{R_{eff}}{R^*})^{4/3} = (\frac{f_{cp}}{f_{ce}})^{4/3} \tag{65}$$

According to Appendix A, the ratio of $W_6$ to $W_0$ is given as:

$$\frac{W_6}{W_0} = (\frac{f_{cp}}{f_{ce}})^2 (\frac{k_e}{k_{el}})^{-1} \tag{66}$$

Substituting Eq. (66) into Eq. (65) gives:

$$\frac{f_{cp}}{f_{ce}} = (\frac{k_e}{k_{el}})^{3/2} \tag{67}$$

These calculation methods are summarised in Table 4. In the work of Pasha et al. (Pasha et al., 2014), a constant $k_e$ is used to check the sensitivity of $f_{cp}$ to $\alpha_p$. However, in that case, Eq. (64) predicts a constant $f_{cp}$ at large plastic deformation whilst Eq. (67) gives $f_{cp}$=$f_{ce}$. Thus, only the case of varying $k_e$ with $\alpha_{\max}$ is focused here.

The predictions of these methods are compared with the results in Ning (1995) and Du et al. (2007), as shown in Fig. 7, with the particle properties and parameters shown in Tables 1-3. Here, $k_e$ is calculated based on Eq. (46) with $k_{el}$ in Table 3. For the ruthenium particle in Du et al. (2007), $k_p$ shows a decrease at small plastic deformation but the decrement is not provided in their work, thus, $k_p$ is set to be $k_p$=minimum($k_p$, $k_e$) for the calculations shown in Fig. 7 for convenience. For ammonium fluorescein particle, all methods predict a larger $f_{cp}$ than that of the non-linear elasto-plastic and adhesive contact model in Ning et al. (1995). For ruthenium particle, compared to the FEM results in Du et al. (2007), Eq. (64) largely underestimates the $f_{cp}$, but Eqs. (59) and (67) show a good agreement, whilst Eq. (61) overestimates $f_{cp}$ significantly. Meanwhile, Eq. (61) predicts $f_{cp}$ to be less than $f_{ce}$ at small plastic deformation, which is not realistic. Thus, the predictions of Eqs. (59) and (67) are more practical than Eqs. (61) and (64). It should be noted that till now, there is no theory to accurately calculate $f_{cp}$, these methods require further validation by various FEM simulations or experimental investigations.

Table 4. Summary of the calculation methods of maximum pull-off force.

Fig. 7. Comparisons of maximum pull-off force $f_{cp}$ predicted by different methods with (a) Ning (1995) and (b) Du et al. (2007) under different maximum normal overlaps $\alpha_{\max}$ (i.e. the overlap at which the unloading commences).

### 3.5 Time step

The critical time step can be estimated from the single degree-of-freedom system of a mass $m$ connected to ground by a spring of stiffness $k_e$, for which the critical time step $\Delta T_{\text{crit}}$ is given as:

$$\Delta T_{crit} = 2\sqrt{\frac{m}{k_e}} \tag{68}$$

As shown in section 3.3, $k_e$ in the unloading process is less than the ones predicted by Hertz model:

$$k_e \le 2ER\sqrt{\alpha_{\max} / R} \tag{69}$$

Thus, the critical time step could be given as:

$$\Delta T_{crit} = \pi R\sqrt{\frac{\rho}{E}}\sqrt{\frac{8}{3\pi}}\left(\frac{\alpha_{\max}}{R}\right)^{-1/4} \tag{70}$$

The critical time step based on Rayleigh surface wave is given as (Raji, 1999):

$$\Delta T_R = \frac{\pi R}{0.1631\upsilon + 0.8766}\sqrt{\frac{\rho}{G}} \tag{71}$$

By comparing Eqs. (70) and (71), given as:

$$\frac{\Delta T_{crit}}{\Delta T_R} = \frac{0.1631\upsilon + 0.8766}{\sqrt{1+\upsilon}}\sqrt{\frac{4}{3\pi}}\left(\frac{\alpha_{\max}}{R}\right)^{-1/4} \tag{72}$$

The variation of $\Delta T_{\mathrm{crit}}/\Delta T_R$ with Poisson's ratio $v$ is shown in Fig. 8. For most granular systems, $0.2 \le v \le 0.4$, $\Delta T_{\mathrm{crit}}/\Delta T_R$ is 0.9-1.0 for $\alpha_{max}/R$=0.1. As the maximum normal overlap is usually less than $0.1R$ and $\Delta T_{\mathrm{crit}}/\Delta T_R$ is inversely proportional to $\alpha_{max}/R$, $\Delta T_{\mathrm{crit}}/\Delta T_R$ could be roughly assumed to be 1 (using its lower limit). Therefore, the time step in DEM simulation for the contact model developed in this work could also be evaluated from the Rayleigh time step in Eq. (71).

It should be noted that the above analysis only provides the maximum limit of the time step, as it does not consider the details of the contact process. Thus, the value of time step used in DEM simulation should be evaluated based on the granular system. For most systems, it can be assumed that $\Delta t$=(0.1~0.25)$\Delta T_R$ for a good balance between the computational time and accuracy. For particulate systems with high collision velocity, e.g. impact of particle against a wall, to accurately detect the contact details (e.g. the line EF in Fig. 1), much smaller time step should be used, i.e. $\Delta t$=0.01~0.1$\Delta T_R$.

Fig. 8. Variation of the ratio of critical time step $\Delta T_{crit}$ to Rayleigh time step $\Delta T_R$ with Poisson's ratio.

### 3.6 Summary

In the contact model developed in this work, several parameters are involved, but most of them can be calculated based on the contact properties. For example, $f_y$ from Eq. (28), $\alpha_y$ from Eq. (29), $k_c$ from Eq. (38) or Eq. (39), $k_e$ from Eq. (46) for the contact after yielding, $f_{cp}$ from Eq. (59) . Thus, apart from the particle properties and interaction parameters in conventional contact models (i.e. Hertz-Mindlin model), such as Young's modulus, friction coefficient, etc., there are four additional input parameters: surface energy $\Gamma$, two stiffnesses (i.e. stiffness $k_{el}$ for the contact before yielding, plastic stiffness $k_p$), and yield contact pressure $p_y$. Auto-calculation and estimation of the parameters involved in the model could be found from supplementary excel files.

1) Surface energy $\Gamma$: it can be experimentally measured from the drop test (Zafar et al., 2014) or centrifuge method (Nguyen et al., 2010). However, for particle with large surface energy and small yield contact pressure, the experiment test could lead to large error (i.e. the contact is already yield during the test). In this case, surface energy $\Gamma$ can be roughly estimated from Hamaker constant of material (Marshall and Li, 2014) and then further calibrated by DEM simulation of bulk particle flow.
2) Stiffnesses $k_{el}$ and $k_p$: plastic stiffness $k_p$ is calculated directly from the loading curve, while minimum elastic stiffness $k_{el}$ could be calculated by Eq. (45) with $k_{e0}$ and $\alpha_{\max0}$ extracted from unloading curve, where large loading force is needed to yield the contact in the indentation test. If no experimental data of the loading/unloading curves is available, $k_{el}$ can be estimated from yield contact pressure using Eq. (52), and $k_p$ can be assumed to be the same as $k_{el}$.
3) Yield contact pressure $p_y$: it could be preferentially calculated from the yield stress $\sigma_y$ using Eq. (32) or Eq. (33). If high impact velocity is involved, the sensitivity of yield stress to strain rate should be considered by Eq. (34). The yield stress here is referred to the compressing yield stress instead of the pulling yield stress. Usually, $\sigma_y/E$ is 0.001~0.1 for most materials (Burgoyne and Daraio, 2014; Du et al., 2007; Etsion et al., 2005; Mesarovic and Johnson, 2000). As the state of the contact (yield or not) is strongly determined by the yield contact pressure, $\sigma_y$ should be carefully examined or calibrated.

## 4 Critical sticking velocity

In this section, the normal impact of a spherical particle against a plane wall is simulated, and the effect of yield contact pressure on the critical sticking velocity is investigated. Based on the surface energy $\Gamma$ and yield contact pressure $p_y$, three kinds of characteristic velocity could be obtained from Fig. 4 and Appendix A:

$$V_s = \sqrt{\frac{2W_0}{m^*}} = 1.84\frac{(\Gamma / R)^{5/6}}{\rho^{1/2} E^{*1/3}} \quad (73)$$

$$V_{y0} = \sqrt{\frac{2W_2}{m^*}} = \sqrt{\frac{f_y^{\ 2}}{m^* k_{el}}} = (\frac{\pi}{2E^*})^2 (\frac{2}{5\rho})^{1/2} p_y^{\ 5/2} \quad (74)$$

$$V_y = \sqrt{\frac{2(W_2 - W_1)}{m^*}} = \sqrt{\frac{f_y^{\ 2} - f_0^2}{m^* k_{el}}} \quad (75)$$

where $m=m^*$ is the particle mass; $R=R^*$ is the particle radius; $V_s$ is the adhesive sticking velocity without considering plastic deformation, below which the particle will stick to the wall due to the adhesive work, and $V_s$ here has the same value as predicted by JKR theory; $V_{y0}$ is the yielding velocity without considering cohesion effect, above which the contact will have plastic deformation; $V_y$ is the corresponding yielding velocity when the cohesion effect is included.

If the particle approaches the plane wall, it will gain energy as a result of the attractive force between them (i.e. negative force, $f=-f_0$ to $f=0$ in line BC in Fig. 1). Thus, the total initial energy involved in the impact is given as:

$$W_i = \frac{1}{2} m V_i^2 + W_1 \quad (76)$$

where the effects of potential energy (i.e. gravity) and viscous damping are omitted. When the particle velocity is reduced to zero, part of total initial energy is converted into stored elastic energy, $W_e$, and the remainder is dissipated through plastic deformation, ($W_i$-$W_e$). If the stored elastic energy, $W_e$, is larger than the adhesion work $W_c$ required to separate the particle from the wall, then the particle will rebound. Otherwise, the particle will remain adhered to the wall. The plastic adhesive sticking velocity $V_{ys}$ is defined as the critical impact velocity, at which the bound velocity is zero, i.e. $W_e=W_c$. Depending on the yield contact pressure, at the impact velocity of $V_i=V_{ys}$, the contact could be yielded or not yielded at the end of the loading process. The critical yield contact pressure distinguishing these two cases is governed by either of the following criteria:

$$W_0 = W_2 - W_1 \text{ or } V_y = V_s \quad (77)$$

For the case with $V_y<V_s$, the contact starts yielding before the end of the loading process at the impact velocity of $V_i=V_{ys}$. In this case, the plastic adhesive sticking velocity $V_{ys}$ is strongly affected by the maximum pull-off force $f_{cp}$. According to Appendix A, $W_e$ and $W_c$ at the impact velocity of $V_i$=$V_{ys}$ are given as:

$$W_e = W_4 = \frac{f_{\max}^{\ 2}}{2k_e} = W_c = W_5 + W_6 = \frac{1}{2}\frac{f_{cp}^{\ 2}}{k_e}(1 + \frac{56}{81}\frac{k_e}{k_c}) \qquad (78)$$

where $f_{max}$ is related to the impact kinetic energy:

$$\frac{mV_{ys}^{\ 2}}{2} = -W_1 + W_2 + W_3 + W_4 = \frac{mV_y^{\ 2}}{2} + \frac{f_{\max}^{\ 2} - f_y^{\ 2}}{2k_p} \qquad (79)$$

By initially assuming $f_{max}$=$f_y$, $V_{ys}$ could be obtained using an iterative method: $\alpha_{max}$ from Eq. (10), $k_e$ from Eq. (46), $f_{cp}$ from Eq. (59), $V_{ys}$ from Eq. (79), and then new $f_{max}$ from Eq. (78) for next iteration. For the case with $V_y \geq V_s$, the contact has not yet yielded at the end of the loading process at the impact velocity of $V_i$=$V_{ys}$, thus, $V_{ys}$ is not affected by $f_{cp}$, and it is given by $V_{ys}$=$V_s$.

The variation of plastic adhesive sticking velocity $V_{ys}$ with yield contact pressure $p_y$ is shown in Fig. 9, where viscous damping is not included. The physical properties of the ammonium fluorescein particle and silicon wall are shown in Table 5, which are updated and summarised by Kim and Dunn (2007). The adhesive sticking velocity is $V_s$=0.66 m/s. Instead of using the fitting value ($k_e$, $k_p$) in Table 2, $k_p$ is assumed to be equal to be $k_{el}$, and $k_{el}$ is given by Eq. (52). At small yield contact pressure, $V_{ys}$ is much larger than $V_s$, which is expected for adhesive contact with plastic deformation. With an increase in yield contact pressure, $V_{ys}$ decreases until it reaches the critical point, where $p_y$ meets the criterion of Eq. (77). With a further increase in the yield contact pressure, $V_{ys}$ does not change anymore and remains equal to the adhesive sticking velocity $V_s$.

It should be noted that there is a special case, i.e. $f_y$<$f_0$ or $W_2$<$W_1$ in Fig. 4, as shown in the shadow zone in Fig. 9, in which the attractive force would always induce plastic deformation as long as the particles can be brought into contact. This is known as adhesion-induced plastic deformation or jump-in induced plastic deformation. If $k_{el}$ is given by Eq. (52), the critical yield contact pressure could be derived from $f_y$=$f_0$ or $W_2$=$W_1$, given as:

$$\frac{p_y}{E} = \frac{2}{\pi^{2/3}}(\frac{5}{6})^{1/6}(\frac{\Gamma}{ER})^{1/3} = 0.9(\frac{\Gamma}{ER})^{1/3} \qquad (80)$$

Thus, large surface energy corresponds to larger critical yield contact pressure. A non-dimensional number, cohesion-yield number, is proposed here:

$$\mathrm{CY} = \frac{{p_y}^3 R}{E^2 \Gamma} \tag{81}$$

It is also close to the ratio of $f_{y0}/f_{ce}$, which is 1.1 times of $CY$. For a given material, with a decrease in particle size, $CY$ decreases, and there is a critical particle size, beyond which adhesion-induced plastic deformation could occur. Below this critical particle size, the smaller the particle size, the more cohesive the particle behaves than the one predicted by traditional JKR theory. Namely, particle needs to be more cohesive (i.e. larger surface energy) in JKR theory to get the same critical sticking velocity as predicted by the ones with plastic deformation.

Table 5． Physical properties of particle and wall in the normal impact test.

Fig. 9. Variation of critical velocity with yield contact pressure, where the velocity is normalised by adhesive sticking velocity $V_s$ predicted by JKR theory; $V_{ys}$ is the critical sticking velocity considering both cohesion and plastic deformation; $V_{y0}$ (not considering cohesion effect) and $V_y$ (considering cohesion effect) are the yielding velocity, above which the contact could be yielded.

The plastic adhesive sticking velocity $V_{ys}$ is also examined by comparing the value predicted by the contact model in this work with the experimental data of Wall et al. (1990), as shown in Fig. 10 and Table 6. Four particle radius are used, i.e. $R$=3.445, 2.45, 1.72 and 1.29 μm, with other physical properties shown in Table 5. For the cases without considering viscous damping (i.e. $e_0$=1), $V_{ys}$ is analytically calculated by solving Eqs. (78) and (79). For the cases considering viscous damping, a DEM simulation is carried out, and the corresponding restitution coefficient in Eq. (26) is set to be $e_0$=0.81, which corresponds to the limit restitution coefficient in Wall (1990). Viscous damping leads to larger value of $V_{ys}$, with a relative increment of about 10%, as shown in Table 6. It is clear that considering both viscous damping and plastic deformation in the simulation could lead to a better agreement with the experimental data of Wall et al. (1990). As shown in Table 6, the critical sticking velocity $V_{ys}$ is much larger than the one $V_s$ predicted by JKR theory, and the difference increases with the decrease in particle size.

Fig. 10. Comparisons of critical sticking velocity $V_{ys}$ predicted by this work with the experimental data of Wall. et al (1990).

Table 6. Critical sticking velocity $V_{ys}$ for different particle size.

**5 Flowability of bulk powder**

The flowability of bulk cohesive powder can be characterised by FT4 rheometer, where a twisted blade rotates anti-clockwise while penetrating down into a particle bed, of which the energy expended for the blade motion is recorded. Such energy is referred as the flow energy, i.e. the input mechanical work of the impeller, and it can be used to infer the flowability of the powder. Powder with poor flowability (e.g. very cohesive) usually corresponds to large flow energy. More details regarding this can be found in the work of Pasha et al. (2020) and Nan et al. (2017). To investigate the combined effects of cohesion and plastic deformation on the flowability of bulk powder, the linear elasto-plastic and cohesive contact model developed in this work is used in DEM simulations. Two systems are simulated here: with and without considering plastic deformation. In the former, plastic stiffness $k_p$ is assumed to be equal to $k_{el}$, and $k_{el}$ is given by Eq. (52) with the yield contact pressure of particles shown in Table 7. In the latter, stiffness $k_{el}$ has the same value as that of the former, while the yield contact pressure is artificially set to a very high value to ensure the contact is always below the yield point to avoid any plastic deformation. Meanwhile, two surface energies are used for particle-particle interaction, i.e. 3.5 and 27 mJ/m$^2$, while the surface energy for particle-wall interaction is constant in all cases, as shown in Table 8. Thus, when considering plastic deformation, based on $p_y$=0.5 MPa shown in Table 7, the cohesion-yield number (Eq. (81)) for particle-particle interaction ($R^*=R/2$) is $CY$=4.2 for $\Gamma$=3.5 mJ/m$^2$, and $CY$=0.5 for $\Gamma$=27 mJ/m$^2$, respectively.

In the DEM simulations, the diameter of the blade and vessel (cylinder) is 23.5 mm and 25 mm,

respectively. The particle radius is 0.4-0.6 mm, with an averaged value of 0.5 mm. The total particle number is about 73,500, forming an initial particle bed height of 19 mm. To reduce the computational time, the tip speed of the blade is set to 0.25 m/s, and only the first 10 mm penetration depth is simulated. The corresponding DEM simulation using Hertz model with JKR theory is also conducted as a reference case. Fig. 11 shows the flow energy of cohesive powder in FT4 rheometer, where the flow energy is normalised by the value predicted by the DEM simulation using Hertz-Mindlin model with JKR theory. For the case without considering plastic deformation, the flow energy predicted by the contact model in this work is very close to the ones predicted by the simulations using Hertz-Mindlin model with JKR theory for the two surface energies used. It is obvious that the powder behaves more "cohesively" when considering plastic deformation, especially for the case with larger surface energy.

Table 7. Physical properties of particle and geometry in the flowability measurement by FT4 rheometer.

Table 8. Interaction parameters of particle and geometry in the flowability measurement by FT4 rheometer.

Fig. 11. Flow energy of cohesive powder in FT4 rheometer predicted by the linear elasto-plastic and adhesive contact model: without and with considering plastic deformation; the flow energy is normalised by the ones predicted by the DEM simulation using Hertz-Mindlin model with JKR theory.

## 6 Conclusions

Based on the work of Thornton and Ning (1998) and Pasha et al. (2014), an improved linear model is developed for elasto-plastic and adhesive contact in DEM simulation. This contact model is then applied to the analysis of single particle impact test and the DEM simulation of bulk particle behaviour in FT4 rheometer. The main results from the present study are summarised as follows:

1) A general and mathematic form is proposed for the contact model, with new criteria to estimate the parameters involved in the model, including various stiffnesses ($k_e$, $k_p$, $k_c$), yield point, maximum pull-off force and time step. The adhesive work and yield work are guaranteed to be the same as the non-linear model of Thornton and Ning (1998). The stiffness in the unloading process is scaled to the square root of the maximum overlap at which unloading commences. The maximum pull-off force increases with the plastic deformation, and it can be reduced to the form predicted by JKR theory if the contact is not yielded. The estimation criteria for elastic stiffness and maximum pull-off force are validated by the loading/unloading curves reported in the literature. The time step can be evaluated based on Rayleigh time step.

2) A new criterion is proposed to calculate the plastic adhesive sticking velocity. The sticking velocity of cohesive particle with a small yield contact pressure is larger than the one predicted by JKR theory. The sticking velocity predicted by the contact model developed in this work is also validated by the experimental data in literature. A cohesion-yield number is proposed to describe the extent of adhesion-induced yielding, in which the attractive force would always induce plastic deformation as long as the particles can be brought into contact. For the particle below critical size, the effect of plastic deformation should be considered and the particles behaves more "cohesively" than the ones predicted by JKR theory.

3) The flowability of bulk powder is strongly affected by the plastic deformation, especially for the particles with large surface energy. The bulk powder in FT4 rheometer behaves more "cohesively" if considering plastic deformation.

**Appendix A**

The work of deformation due to normal contact force in each stage in Fig. 4 is derived as follows:

$$W_0 = \frac{1}{2}(\alpha_{ce} - \alpha_{fe})(\frac{5}{9}f_{ce} + f_{ce}) + \frac{1}{2}\alpha_{ce}(\frac{8}{9}f_{ce} + f_{ce}) = \frac{56}{162}\frac{f_{ce}^2}{k_{cl}} + \frac{17}{162}\frac{f_{ce}^2}{k_{el}} \tag{A1}$$

$$W_1 = \frac{f_0^2}{2k_{el}} = \frac{64}{162}\frac{f_{ce}^2}{k_{el}} \tag{A2}$$

$$W_2 = \frac{f_y^2}{2k_{el}} \tag{A3}$$

$$W_3 = \frac{1}{2}(f_y + f_{\max})\cdot(\alpha_{\max} - \alpha_y) - \frac{1}{2}f_{\max}(\alpha_{\max} - \alpha_p) = \frac{(f_{\max}^2 - f_y^2)}{2k_p} - \frac{f_{\max}^2}{2k_e} \tag{A4}$$

$$W_4 = \frac{f_{\max}^{\ 2}}{2k_e} \tag{A5}$$

$$W_5 = \frac{1}{2}\frac{(\frac{8}{9}f_{cp})^2}{k_e} = \frac{64}{162}\frac{f_{cp}^{\ 2}}{k_e} \tag{A6}$$

$$W_6 = \frac{1}{2}(\alpha_{cp} - \alpha_{fp})(\frac{5}{9}f_{cp} + f_{cp}) + \frac{1}{2}(\alpha_{c0} - \alpha_{cp})(\frac{8}{9}f_{cp} + f_{cp}) = \frac{56}{162}\frac{f_{cp}^{\ 2}}{k_c} + \frac{17}{162}\frac{f_{cp}^{\ 2}}{k_e} \tag{A7}$$

**Appendix B**

In the Hertz model with JKR theory, the contact force and normal stiffness are given as:

$$f_{H-JKR} = \frac{4E^* a^3}{3R^*} - \sqrt{8\pi\Gamma E^*}\, a^{3/2} \tag{B1}$$

$$k_{H-JKR} = 2E^* a \frac{\sqrt{f_H / f_{ce}} - 1}{\sqrt{f_H / f_{ce}} - 1/3} \tag{B2}$$

where $f_{ce}=1.5\pi\Gamma R^*$; $a$ is the contact radius; $f_H$ is the equivalent Hertz force with the same contact radius. $a$ and $f_H$ are given as:

$$\alpha = \frac{a^2}{R^*} - \left(\frac{2\pi\Gamma a}{E^*}\right)^{1/2} \tag{B3}$$

$$f_H = \frac{4E^* a^3}{3R^*} \tag{B4}$$

where $\alpha$ is the normal physical overlap. At the point $\alpha=0$, the contact radius is given as:

$$a = (2\pi\frac{\Gamma R^{*2}}{E^*})^{1/3} \tag{B5}$$

Substituting Eq. (B5) into Eqs. (B4) and (B2), given as:

$$f_H / f_{ce} = \frac{16}{9} \tag{B6}$$

$$k_{H-JKR,\,\alpha=0} = (\frac{16}{27}\pi)^{1/3}(\Gamma E^{*2} R^{*2})^{1/3} \tag{B7}$$

Similarly, at the point $\alpha=\alpha_0$, the contact force $f_{H-JKR}=0$, given as:

$$a = (\frac{9\pi}{2}\frac{\Gamma R^{*2}}{E^*})^{1/3} \tag{B8}$$

$$f_H / f_{ce} = 4 \tag{B9}$$

$$k_{H-JKR,\,\alpha=\alpha_0} = \frac{6}{5}(\frac{9}{2})^{1/3}\pi^{1/3}(\Gamma E^{*2} R^{*2})^{1/3} \tag{B10}$$

$$\alpha_0 = (\frac{3}{4})^{1/3}(\frac{\pi^2 R^* \Gamma^2}{E^{*2}})^{1/3} \tag{B11}$$

At this point, the normal stiffness of Hertz model with the same overlap is given as:

$$k_{H,\ \alpha=\alpha_0} = 2E^*\sqrt{R^*\alpha_0} = 2(\frac{3}{4})^{1/6}\pi^{1/3}(\Gamma E^{*2}R^{*2})^{1/3} \quad \text{(B12)}$$

By comparing (B12) and (B10), given as:

$$\frac{k_{H-JKR,\ \alpha=\alpha_0}}{k_{H,\ \alpha=\alpha_0}} = \frac{3\sqrt{3}}{5} = 1.04 \quad \text{(B13)}$$

Thus, for the normal overlap larger than $\alpha_0$, i.e. $f_{H\text{-}JKR}>0$, the normal stiffness in Hertz model with JKR theory (Eq. (B2)), could be estimated by the ones predicted by Hertz model (Eq. (41)) at the same normal overlap, which is also clear illustrated in Fig. 12.

Fig. 12. Variation of normal stiffness with normal overlap in Hertz model and Hertz-JKR model for the case with $R^*$=2.45 μm, $E^*$=1.3 GPa and $\Gamma$=0.2 J/m$^2$.

## Acknowledgments

The first author is grateful to the National Natural Science Foundation of China (Grant No. 51806099). The authors are also thankful to DEM Solutions, Edinburgh, UK, for providing a special license for the EDEM software for use in this work. The first author is also thankful to Professor Mojtaba Ghadiri, University of Leeds, UK, for the inspiration and encouragement on this work.

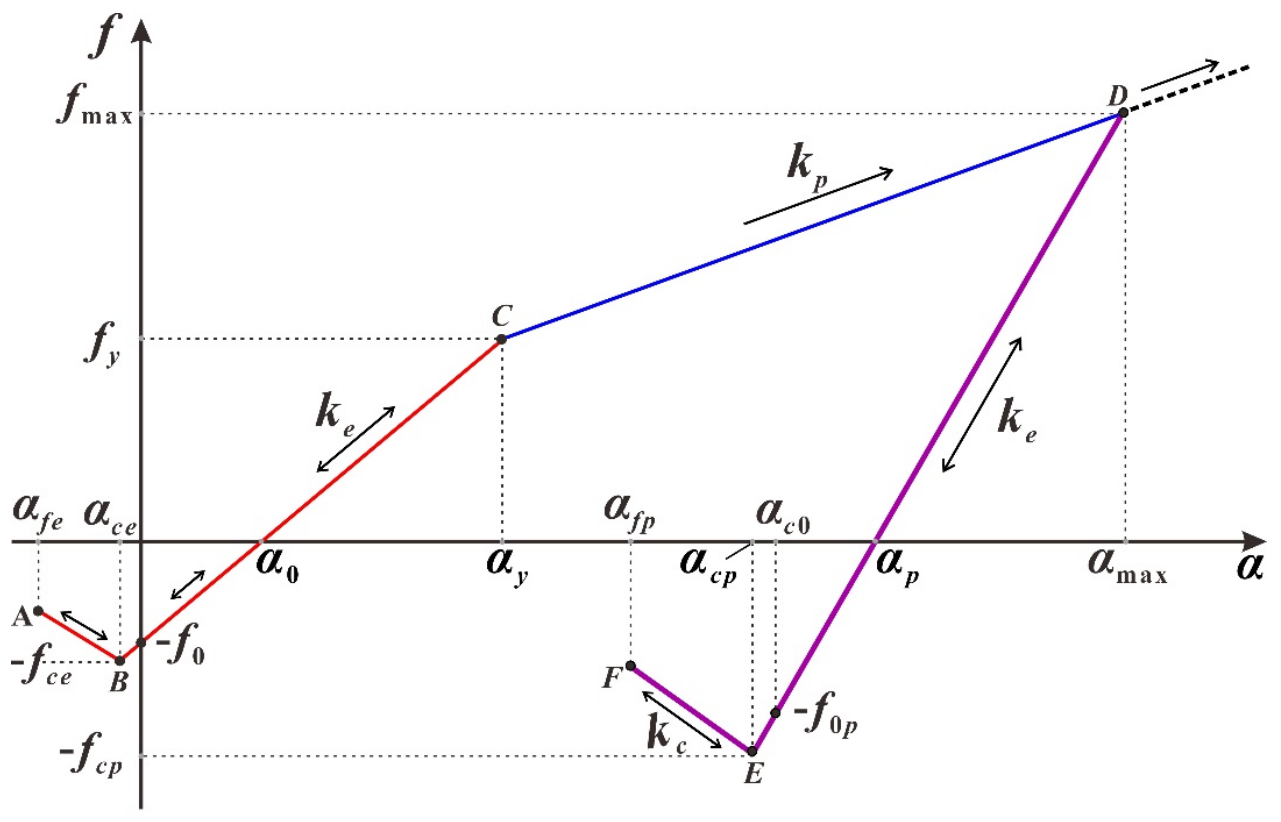

Fig. 1. Schematic diagram of the normal force *f*-overlap *α* relationship in the improved linear elasto-plastic and adhesive model.

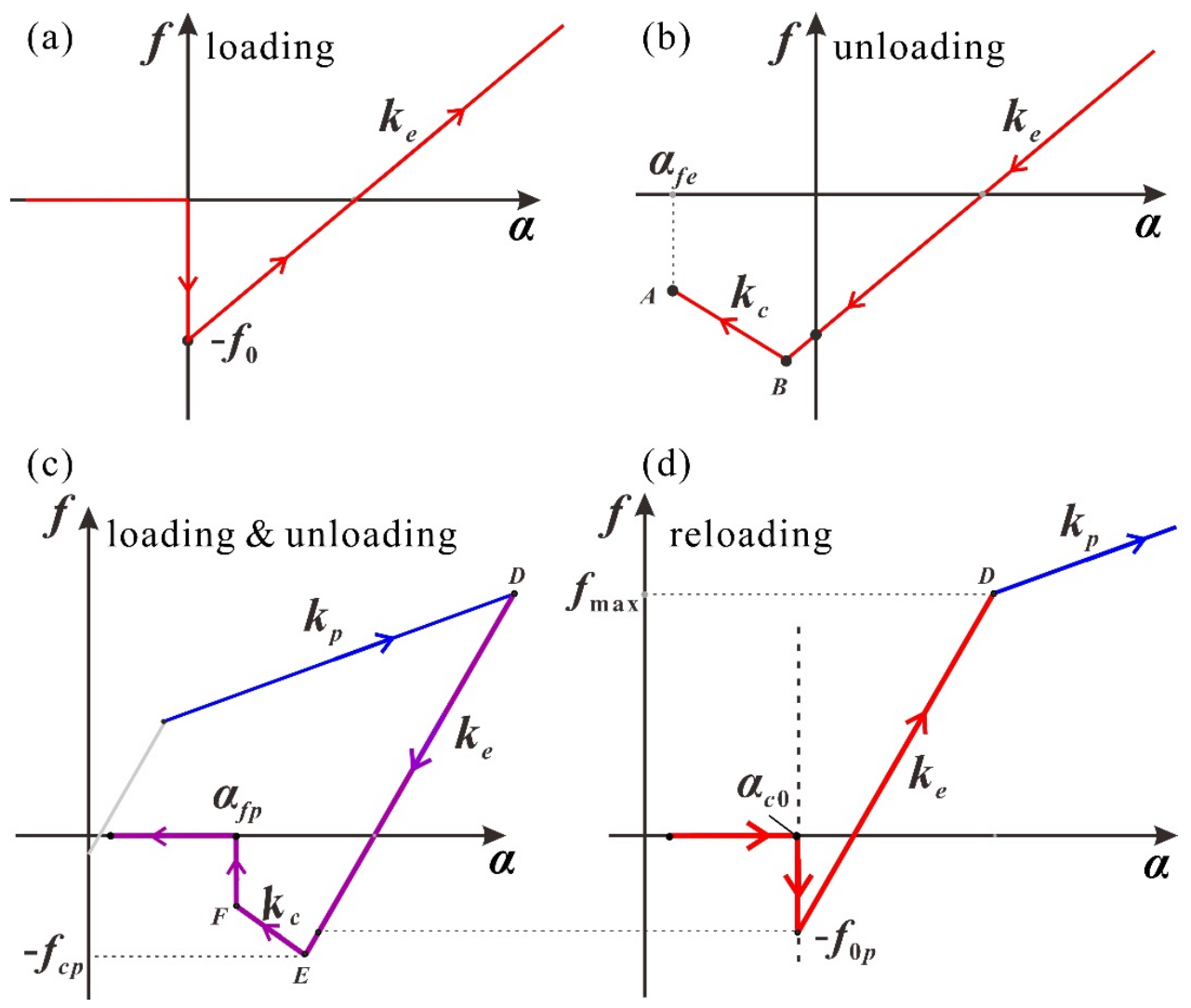

Fig. 2. Schematic diagram of the normal force $f$-overlap $\alpha$ relationship in the loading/unloading and reloading processes for the contact: (a-b)-without plastic deformation ($f_{\max}<f_y$); (c-d)-with plastic deformation ($f_{\max}>f_y$).

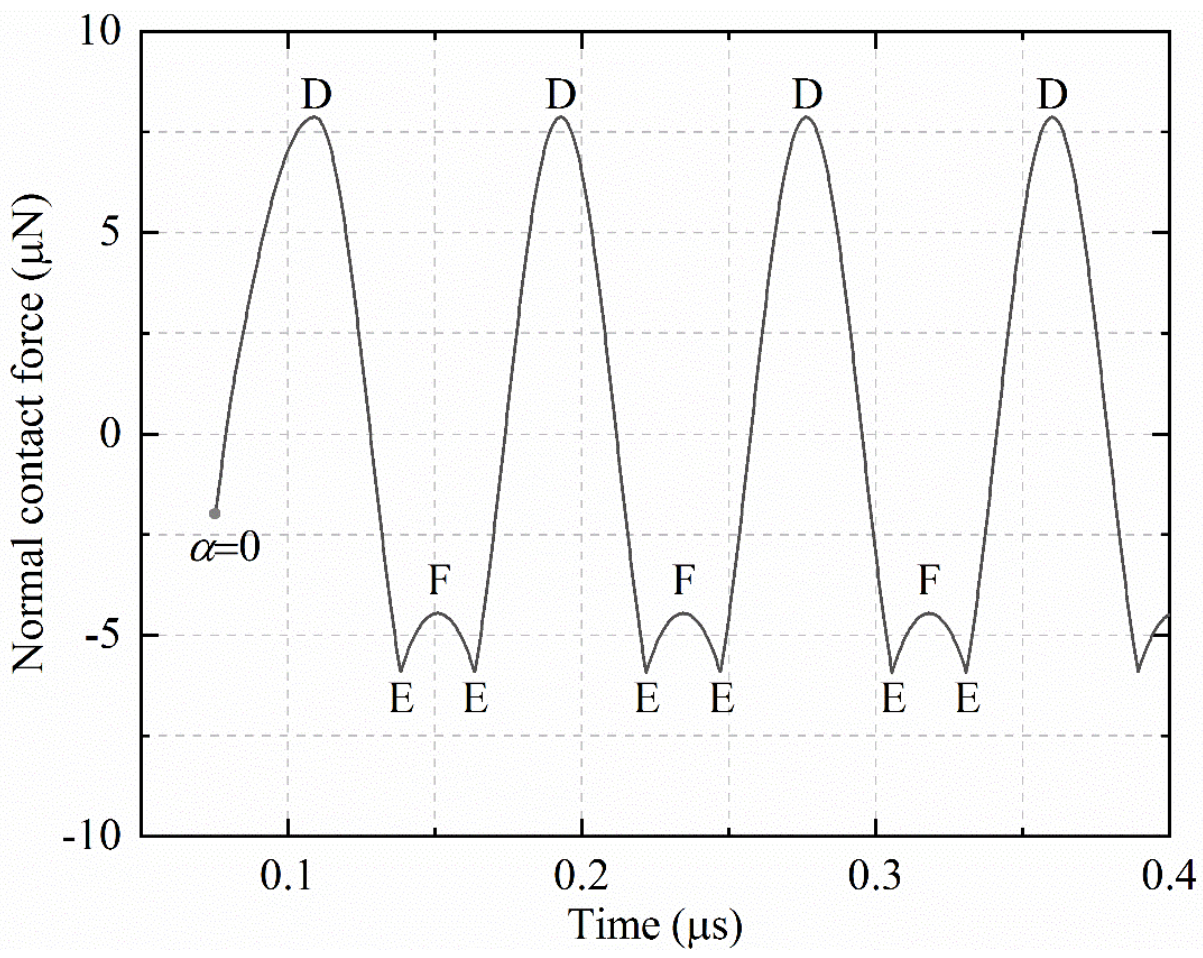

Fig. 3. Oscillation of normal contact force with time if there is no viscous damping after yielding.

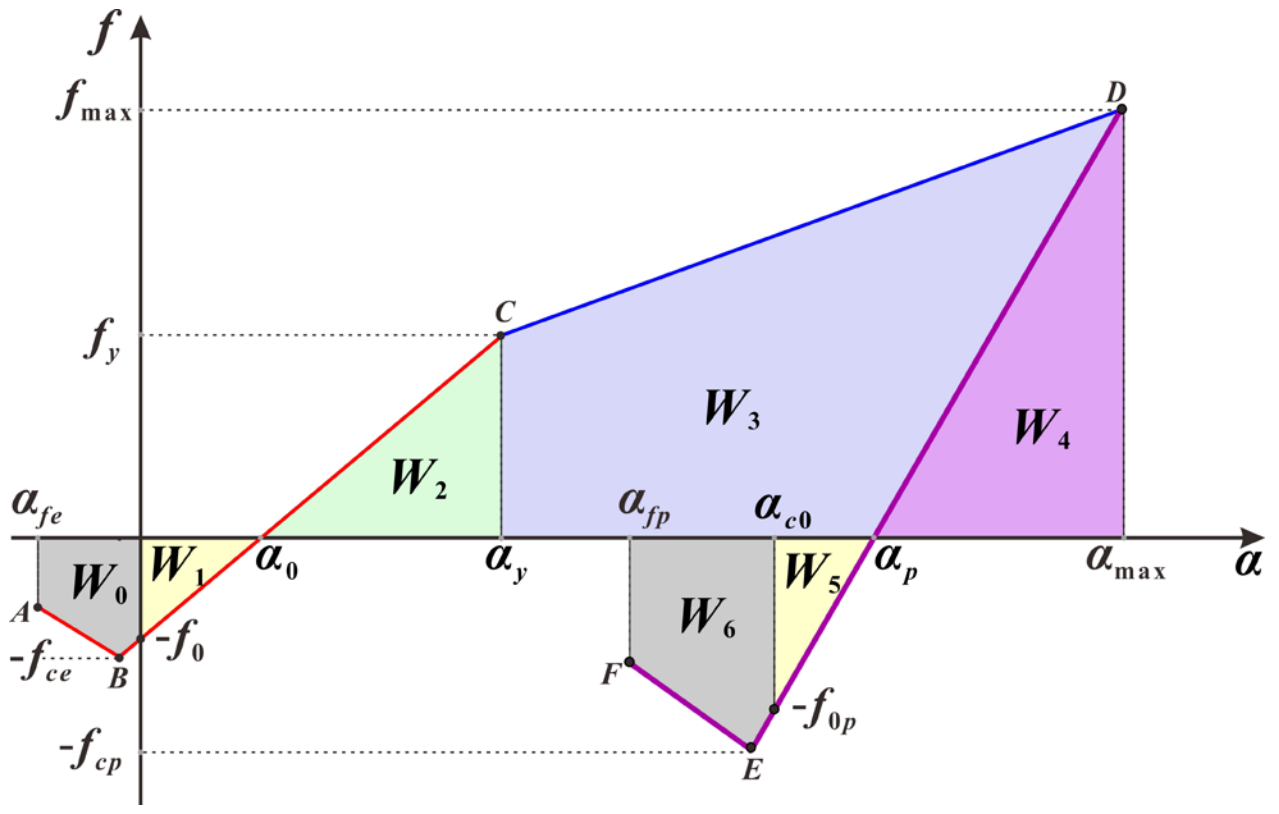

Fig. 4. Work of deformation due to normal contact force in different contact stages.

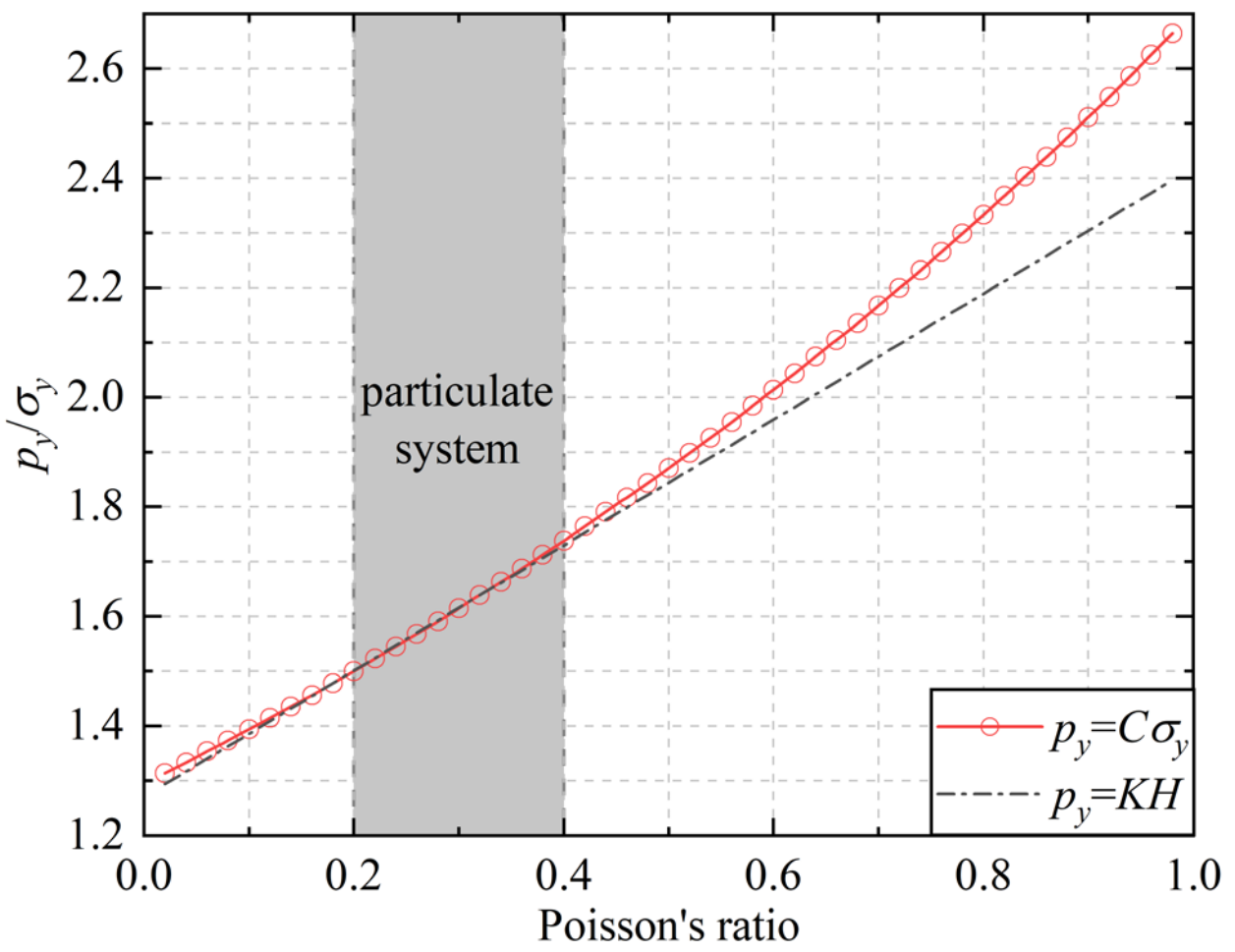

Fig. 5. Variation of the ratio of yield contact pressure $p_y$ to yield stress $\sigma_y$ with Poisson's ratio.

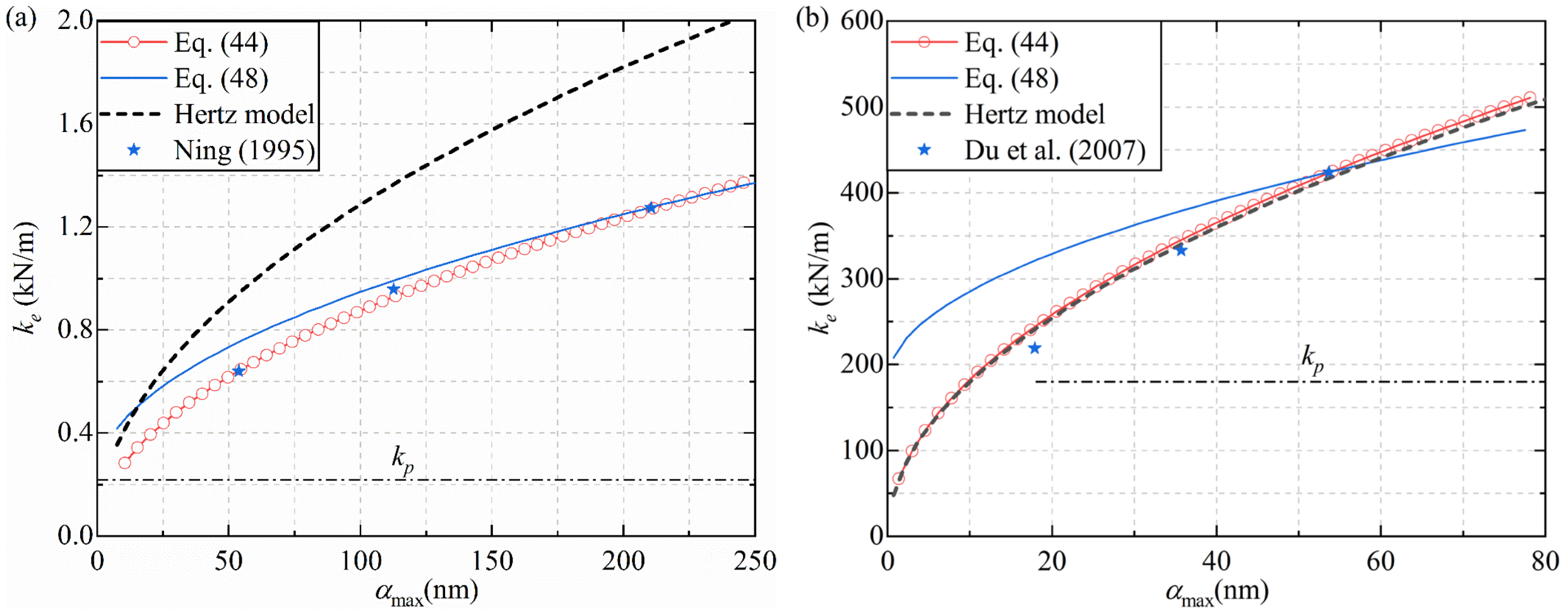

Fig. 6. Comparisons of stiffness $k_e$ predicted by different methods with (a) Ning (1995) and (b) Du et al. (2007) under different maximum normal overlaps $\alpha_{max}$ (i.e. the overlap at which the unloading commences).

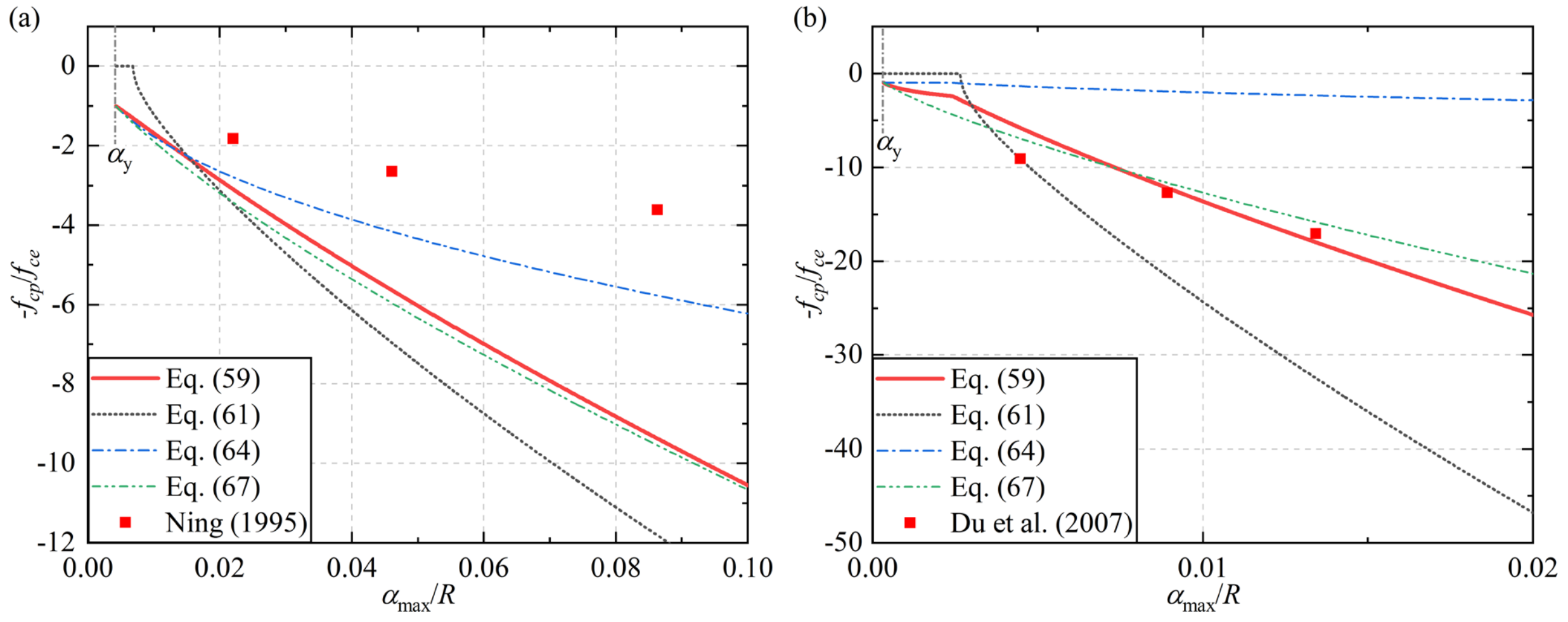

Fig. 7. Comparisons of maximum pull-off force $f_{cp}$ predicted by different methods with (a) Ning (1995) and (b) Du et al. (2007) under different maximum normal overlaps $\alpha_{max}$ (i.e. the overlap at which the unloading commences).

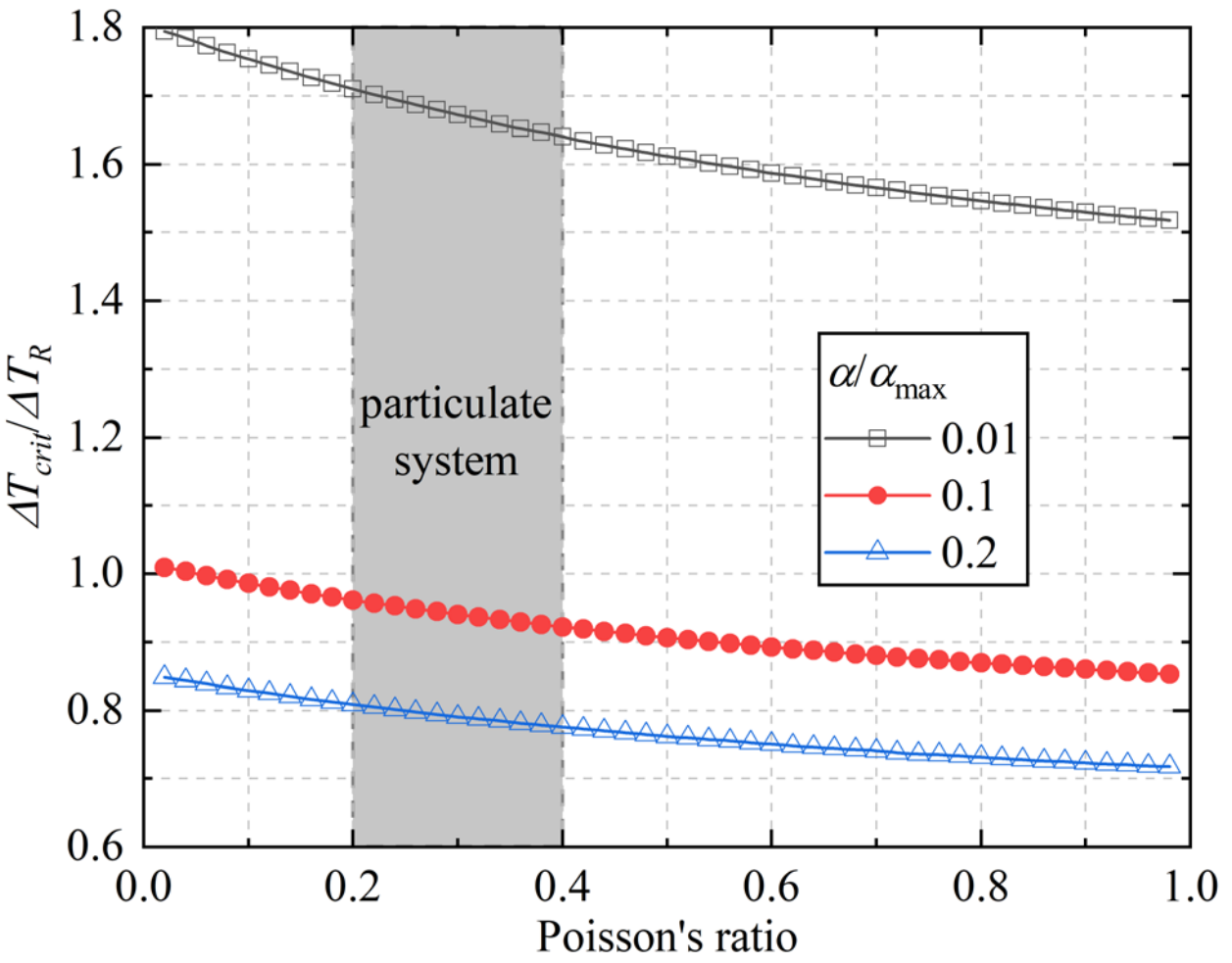

Fig. 8. Variation of the ratio of critical time step $\Delta T_{\mathrm{crit}}$ to Rayleigh time step $\Delta T_R$ with Poisson's ratio.

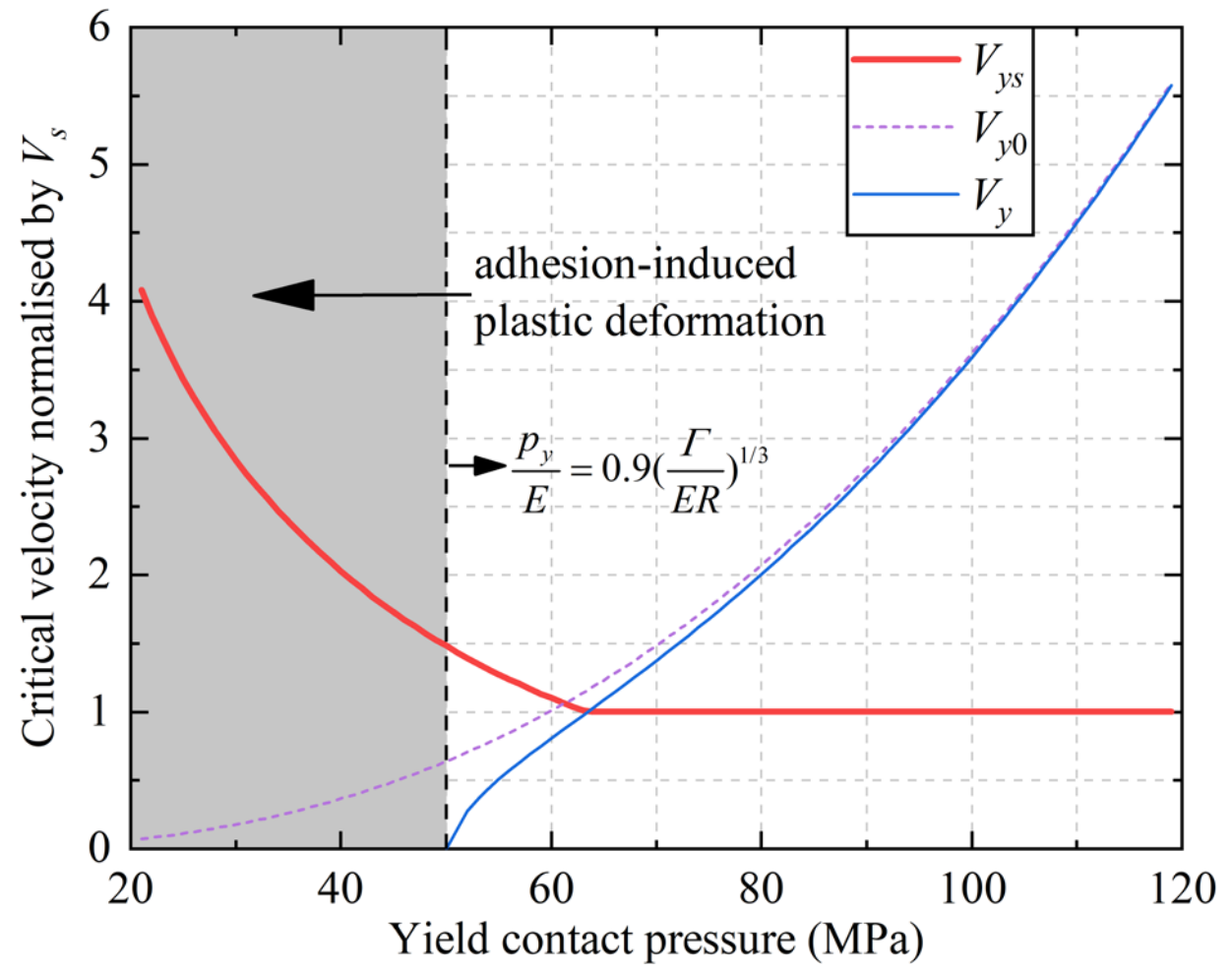

Fig. 9. Variation of critical velocity with yield contact pressure, where the velocity is normalised by adhesive sticking velocity $V_s$ predicted by JKR theory; $V_{ys}$ is the critical sticking velocity considering both cohesion and plastic deformation; $V_{y0}$ (not considering cohesion effect) and $V_y$ (considering cohesion effect) are the yielding velocity, above which the contact could be yielded.

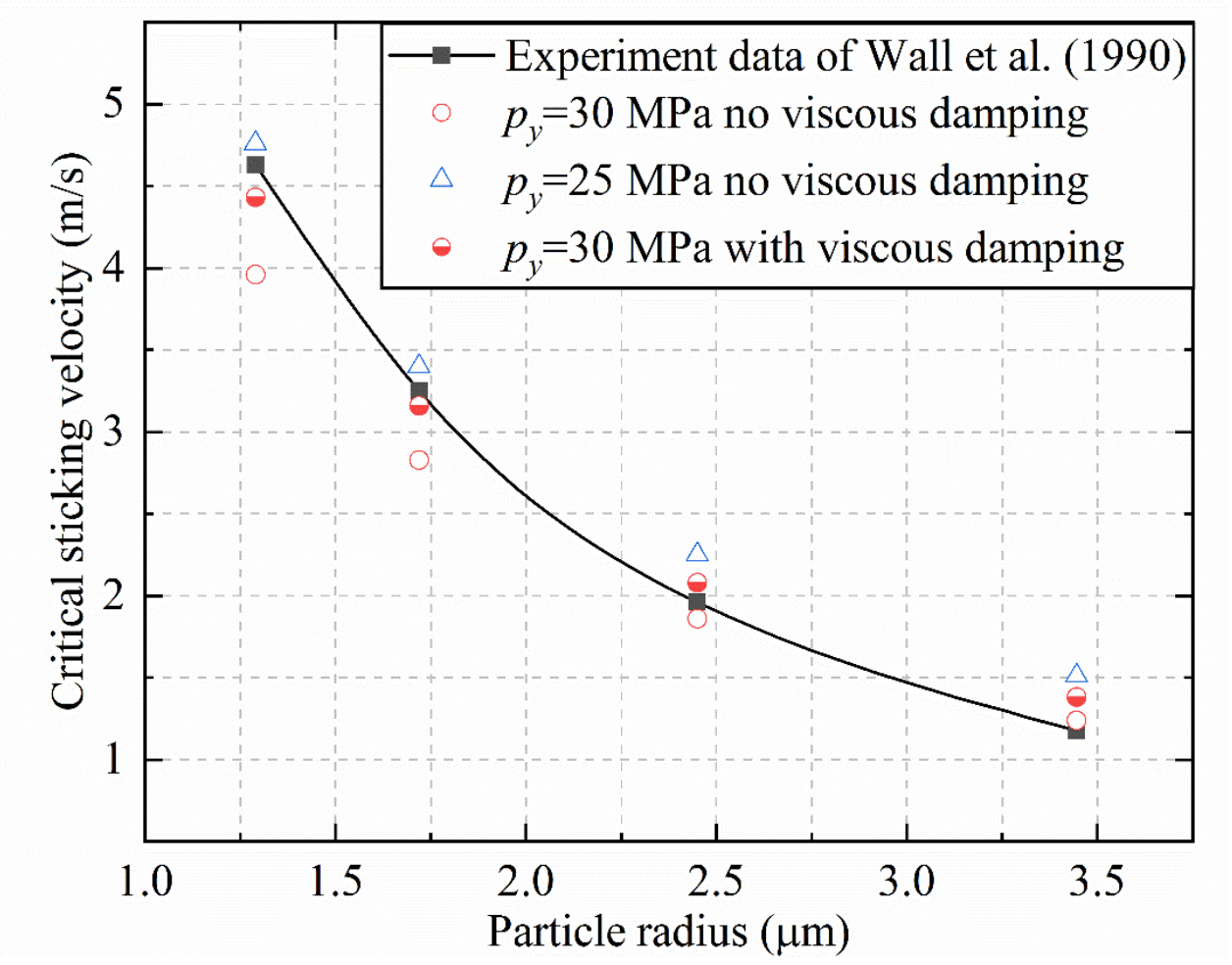

Fig. 10. Comparisons of critical sticking velocity $V_{ys}$ predicted by this work with the experimental data of Wall. et al. (1990).

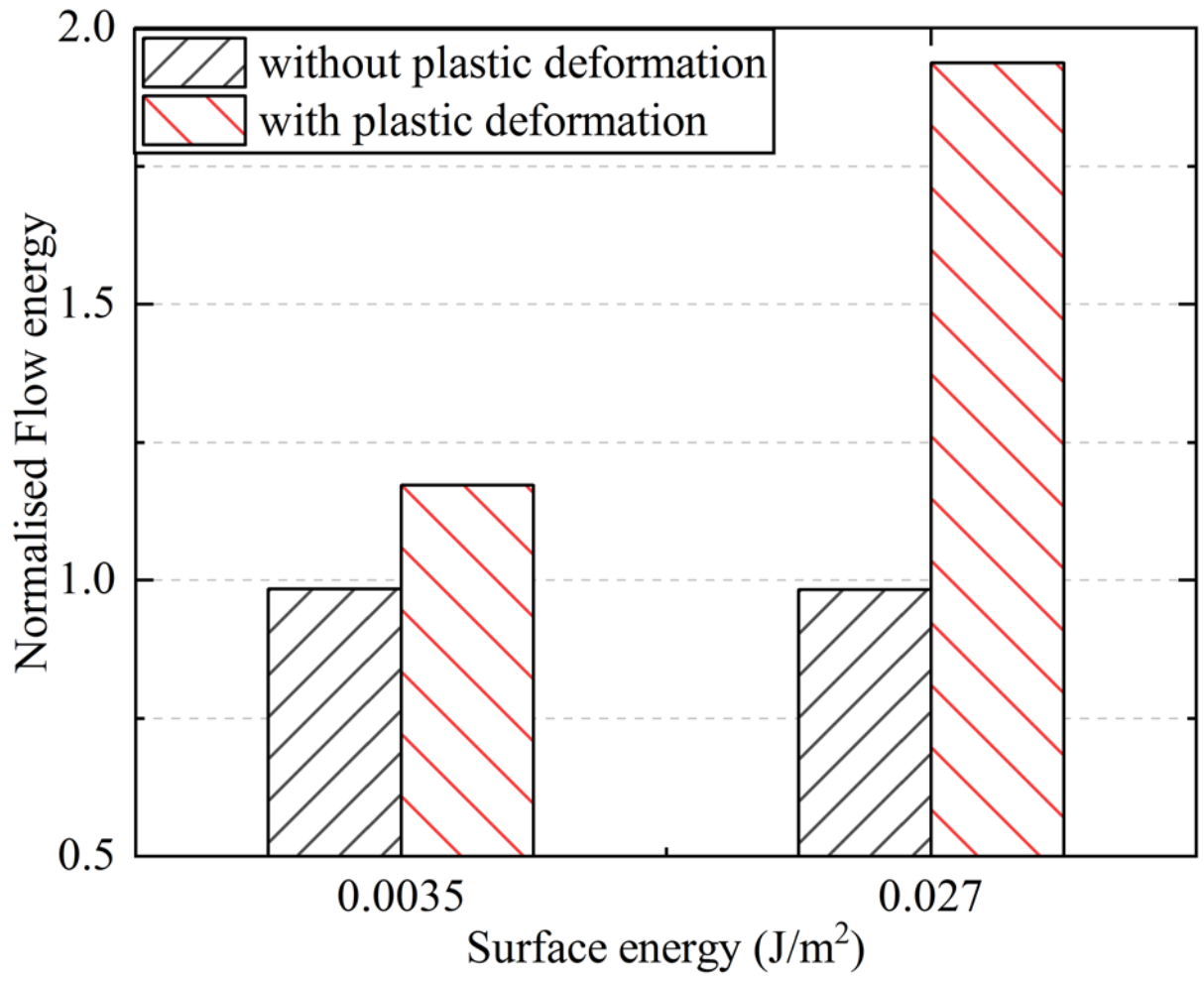

Fig. 11. Flow energy of cohesive powder in FT4 rheometer predicted by the linear elasto-plastic and adhesive contact model: without and with considering plastic deformation; the flow energy is normalised by the ones predicted by the DEM simulation using Hertz-Mindlin model with JKR theory.

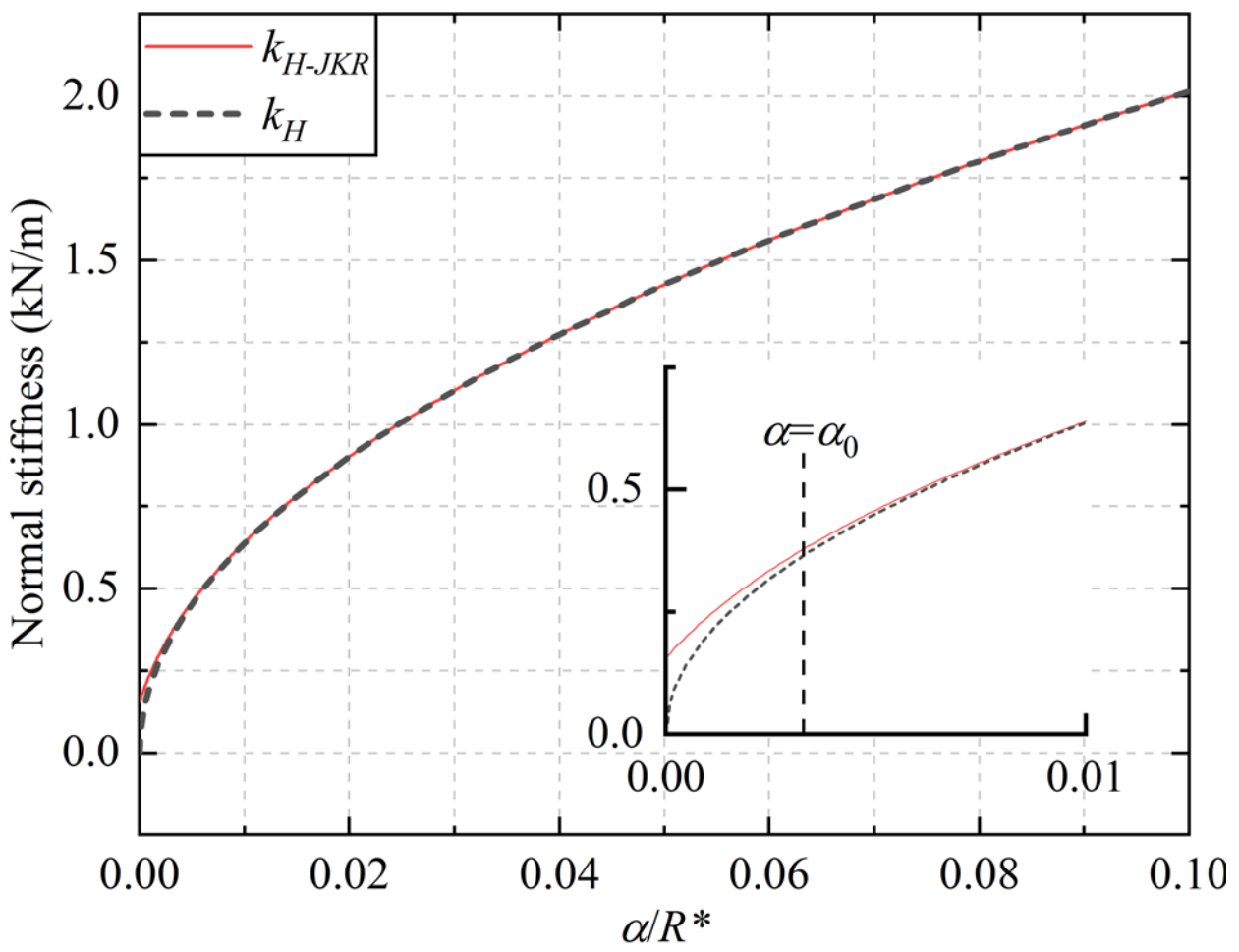

Fig. 12. Variation of normal stiffness with normal overlap in Hertz model and Hertz-JKR model for the case with $R^*$=2.45 μm, $E^*$=1.3 GPa and $\Gamma$=0.2 J/m$^2$.

Table 1. Particle properties used in the work of Ning (1995) and Du et al. (2007).

| Properties | Ning | Du et al. |
|---|---|---|
| Radius (μm), $R$ | 2.45 | 4 |
| Density (kg/m$^3$), $\rho$ | 1350 | - |
| Young's modulus (GPa), $E$ | 1.2 | 410 |
| Poisson's ratio, $v$ | 0.3 | 0.3 |
| Surface energy (J/m$^2$), $\Gamma$ | 0.2 | 1.0 |
| Yield pressure (MPa), $p_y$ | 35.3 | $5.52\times10^3$ * |
| $\pi R^* p_y$ (N/m) | 272 | $6.94\times10^4$ |

* calculated from the yield stress ($p_y=C\sigma_y$)

Table 2. Re-calculated particle contact parameters based on the plot in Ning (1995) and Du et al. (2007).

| Parameters | Ning | Du et al. |
|---|---|---|
| | 42.4 | 8584.4 |
| $f_{max}$ (μN) | 20.9 | 5170.7 |
| | 8.3 | 2132.2 |
| | 210.5 | 53.7 |
| $\alpha_{max}$ (nm) | 112.6 | 35.7 |
| | 53.9 | 17.9 |
| | 177.3 | 33.4 |
| $\alpha_p$ (nm) | 91.1 | 20.2 |
| | 41.3 | 8.2 |
| | -8.4 | 321.5 |
| $f_{cp}$ (μN) | -6.1 | 239.7 |
| | -4.6 | 170.8 |
| | 1274 | $4.2\times10^5$ |
| $k_e$ (N/m) | 958 | $3.3\times10^5$ |
| | 639 | $2.2\times10^5$ |
| $k_p$ (N/m) | 217 | $1.8\times10^5$ |

Table 3. Particle contact parameters at the yield point ($\alpha$=$\alpha_y$) for the particles used in Ning (1995) and Du et al. (2007).

| | Parameters | $k_e = k_{e0}\sqrt{\frac{\alpha_{max}}{\alpha_{max0}}}$ | $k_e = k_p + (k_{e0} - k_p)\sqrt{\frac{\alpha_{max}}{\alpha_{max0}}}$ |
|---|---|---|---|
| Ning | $k_{el}$, N/m | 283 | 418 |
| | $k_H$ ($\alpha$=$\alpha_y$), N/m | 416 | 353 |
| | $\alpha_y$, nm | 10.4 | 7.5 |
| | $\alpha_0$, nm | 7.3 | 4.9 |
| | $k_{el}/k_{cl}$ | 1.8 | 2.7 |
| Du et al. | $k_{el}$, N/m | $6.7\times10^4$ | $2.1\times10^5$ |
| | $\alpha_y$, nm | 1.35 | 0.71 |
| | $\alpha_0$, nm | 0.25 | 0.08 |
| | $k_{el}/k_{cl}$ | 3.9 | 12.6 |

Table 4. Summary of the calculation methods of maximum pull-off force.

| Formulations | Equation number | Remarks |
|---|---|---|
| $\frac{f_{cp}}{f_{ce}}=\frac{-A+\sqrt{A^2+4\frac{k_e}{k_{el}}(\frac{\alpha_p}{\alpha_0}A+1)}}{2}$ | (59) | $A=\frac{\frac{16}{27}}{\frac{56}{162}\frac{k_e}{k_c}+\frac{17}{162}}$ |
| $\frac{f_{cp}}{f_{ce}}=\sqrt{0.7\frac{k_e}{k_{el}}\frac{(\alpha_p-\alpha_y)}{\alpha_0}(2-\frac{\alpha_p-\alpha_y}{R^*})}$ | (61) | re-derived based on Pasha et al. (2014) |
| $\frac{f_{cp}}{f_{ce}}=\frac{f_y+k_e(\alpha_{\max}-\alpha_y)}{f_y+k_p(\alpha_{\max}-\alpha_y)}$ | (64) | not suitable for $k_e$ not varying $\alpha_{\max}$ |
| $\frac{f_{cp}}{f_{ce}}=(\frac{k_e}{k_{el}})^{3/2}$ | (67) | not suitable for $k_e$ not varying $\alpha_{\max}$ |

Table 5． Physical properties of particle and wall in the normal impact test.

| Properties | Particle | Wall |
|---|---|---|
| Radius (μm), $R$ | 2.45 | - |
| Density (kg/m$^3$), $\rho$ | 1350 | 2330 |
| Young's modulus (GPa), $E$ | 1.2 | 166 |
| Poisson's ratio, $v$ | 0.33 | 0.28 |
| Yield pressure (MPa), $p_y$ | - | 120 |
| Surface energy (J/m$^2$), $\Gamma$ | 0.24 | |

Table 6. Critical sticking velocity $V_{ys}$ for different particle size.

| Particle properties | Radius (μm) | 3.445 | 2.45 | 1.72 | 1.29 | Mean relative error |
|---|---|---|---|---|---|---|
| | $V_s$ (m/s) | 0.49 | 0.66 | 0.88 | 1.12 | |
| $V_{ys}$ | Wall. et al (1990) | 1.18 | 1.96 | 3.25 | 4.63 | - |
| | $p_y$ =25 MPa, $e_0$=1 | 1.51 | 2.25 | 3.4 | 4.76 | 12.5% |
| | $p_y$=30 MPa, $e_0$=1 | 1.24 | 1.86 | 2.83 | 3.96 | 9.4% |
| | $p_y$ =30 MPa, $e_0$=0.81 | 1.38 | 2.08 | 3.16 | 4.43 | 7.5% |

Table 7. Physical properties of particle and geometry in the flowability measurement by FT4 rheometer.

| Properties | Particle | Blade | Vessel |
|---|---|---|---|
| Radius (mm), $R$ | 0.2-0.3 | - | - |
| Density (kg/m$^3$), $\rho$ | 2500 | 7800 | 2500 |
| Young's modulus (MPa), $E$ | 63 | 210 | 63 |
| Poisson's ratio, $v$ | 0.2 | 0.3 | 0.2 |
| Contact yield pressure (MPa), $p_y$ | 0.5 | - | - |

Table 8. Interaction parameters of particle and geometry in the flowability measurement by FT4 rheometer.

| Interaction parameters | Particle-Particle | Particle-Blade | Particle-Vessel |
|---|---|---|---|
| Restitution coefficient, $e_0$ | 0.93 | 0.95 | 0.93 |
| Sliding friction, $\rho$ | 0.5 | 0.3 | 0.5 |
| Surface energy (mJ/m$^2$), $\Gamma$ | 3.5, 27 | 3.1 | 2.6 |